\documentstyle[12pt,epsf]{article}

\def\la{\mathrel{\mathpalette\fun <}}

\def\fun#1#2{\lower3.6pt\vbox{\baselineskip0pt\lineskip.9pt
  \ialign{$\mathsurround=0pt#1\hfil##\hfil$\crcr#2\crcr\sim\crcr}}}
\relax
\makeatletter
\@addtoreset{equation}{section}
\makeatother

\let\a=\alpha

     \let\la=\label
   
 \def\nn{\nonumber} \def\bd{\begin{document}}
\def\ed{\end{document}} \def\ds{\documentstyle} \let\fr=\frac
\let\bl=\bigl \let\br=\bigr \let\Br=\Bigr \let\Bl=\Bigl \let\bm=\bibitem
\let\na=\nabla \let\pa=\partial \let\ov=\overline
\newcommand{\be}{\begin{equation}} \newcommand{\ee}{\end{equation}}
\def\ba{\begin{array}} \def\ea{\end{array}}
\newcommand{\bea}{\begin{eqnarray}} \newcommand{\eea}{\end{eqnarray}}
\newcommand{\ra}{\rightarrow} \newcommand{\oh}{\frac{1}{2}}
\newcommand{\lra}{\longrightarrow} \newcommand{\Lra}{\Leftrightarrow}
\newcommand{\ap}{\alpha^\prime} \newcommand{\bp}{\tilde \beta^\prime}
\newcommand{\tr}{{\rm tr} } \newcommand{\sg}{\sqrt {-g}}
\newcommand{\rn}{r_0} \newcommand{\rpl}{r_p} \newcommand{\hoch}[1]{$\,
^{#1}$} \newcommand{\srt}{\sqrt{3}}
\newcommand{\oosrt}{\frac{1}{\sqrt{3}}} \newcommand{\rmi}{r_-}
\newcommand{\Dp}{\Delta_{+}} \newcommand{\Dm}{\Delta_{-}}
\newcommand{\Tr}{{\rm Tr} } \newcommand{\NP}{Nucl. Phys. }
\newcommand{\lag}{{\cal L}} 

\newcommand{\dt}{\tilde{d}}
\newcommand{\ttt}{\hspace{0.7mm}\tilde{\hspace{-1mm}\tilde t}\hspace{0.5mm}}
\newcommand{\mtt}{\hspace{0.7mm}\tilde{\hspace{-0.0mm}\tilde m}\hspace{0.5mm}}
\newcommand{\gtt}{\hspace{0.3mm}\tilde{\hspace{-0.2mm}\tilde g}\hspace{0.5mm}}
\newcommand{\Gtt}{\hspace{0.3mm}\tilde{\hspace{-0.7mm}\tilde G}\hspace{0.5mm}}
\newcommand{\ratio}{\frac{d\tilde{d}}{d+\dt}}
\newcommand{\vaa}{\vec{a_{\a}}}
\newcommand{\vp}{\vec{\phi}}
\newcommand{\Nu}{{\cal V}}
\advance\textheight by \topskip
\begin{document}
\begin{titlepage}

\vspace*{-1.5truecm}

\rightline{IASSNS-HEP 97/28}
\rightline{SLAC-PUB-7468}
\rightline{SNUTP 97-043}
\rightline{SU-ITP 97/13}
\rightline{\tt hep-th/9704142}
\begin{center}
{\Large\bf  CRITICAL  POINTS  AND  PHASE TRANSITIONS \\
\vskip 0.75 cm
IN 5D COMPACTIFICATIONS  OF M-THEORY}
\vskip 0.5 cm
{\large {\bf
Aaron Chou${}^{a}$,
R. Kallosh${}^b$,
J. Rahmfeld${}^b$,\\
\vskip 0.3cm
Soo-Jong Rey ${}^{abcd}$,
M. Shmakova${}^{be}$, 
Wing  Kai Wong${}^{b}$
\footnote{\tt
aschou@slac.stanford.edu,
kallosh@physics.stanford.edu,
rahmfeld@leland.stanford.edu,
sjrey@sns.ias.edu,
shmakova@slac.stanford.edu,
wkwong@leland.stanford.edu}
}}\\
\vskip 0.75 cm
{\it ${}^a$Stanford Linear Accelerator Center, Stanford University,
Stanford, CA 94309\\
 ${}^b$Physics Department, Stanford University, Stanford, CA
94305-4060\\
${}^c$School of Natural Sciences, Institute for Advanced Study,
Princeton, NJ 08540\\
${}^d$Physics Department, Seoul National University, Seoul 151-742
KOREA\\
${}^e$University of Tennessee, Knoxville, TN 37996\\
}

\end{center}
\begin{abstract}
We study critical points of the  BPS mass $Z$, the BPS
string tension $Z_m$, the black hole
potential $V$ and the gauged central charge potential $P$ for
M-theory compactified on Calabi-Yau three-folds. We  first
show that the stabilization equations for $Z$ (determining the 
black hole entropy) take an extremely simple 
form in five dimensions
as opposed to  four dimensions. The stabilization equations for
 $Z_m$ are also very simple and
determine the size of the infinite $adS_3$-throat of the string.
The black hole potential
in general exhibits two classes of critical points:
supersymmetric  critical points which coincide with those of
the central charge and non-supersymmetric critical points.
We then generalize the discussion to  the entire
extended K\"ahler cone
encompassing topologically different but birationally 
equivalent Calabi-Yau three-folds that are connected via flop 
transitions. We  examine  
behavior of the four potentials to probe the
nature of these phase transitions.
We find that $V$ and $P$ are continuous but not smooth
across the flop transition, while 
$Z$ and its first two derivatives, as well
as $Z_m$ and its first derivative, are continuous.
This in turn implies that supersymmetric stabilization of
$Z$ and $Z_m$   for a given 
configuration
takes place in at most one point throughout the entire extended
K\"ahler cone.  
The corresponding black holes (or string states) 
interpolate between different
Calabi-Yau  three-folds. 
At the boundaries of the extended K\"ahler cone we
observe
that electric states become massless and/or magnetic strings become
tensionless.

\end{abstract}
\end{titlepage}

\section{Introduction}
 
In the moduli space encompassing all known perturbative string
theories, it has been discovered~\cite{Wittenvarious}
that there exists a special point at which  perturbative
string theories are strongly coupled and are best described by a more 
fundamental eleven-dimensional theory. For low-energy excitations,
this so-called M-theory is  given by eleven-dimensional
$N=1$ supergravity~\cite{Cremmer}
 appended  with gravitational anomaly cancellation terms.
Not only providing a more fundamental description of non-perturbative 
string theories,  M-theory is also indispensable for constructing 
a phenomenologically viable four-dimensional string compactifications.
Witten~\cite{Wittenstrong} has shown that strongly coupled phase of
$E_8 \times E_8$ heterotic string theory, which is most appropriately
described by M-theory compactified on an orbifold ${\bf S}_1/{\bf Z}_2$, 
is crucial for unification of all gauge and gravitational couplings
at the grand-unification scale~\footnote{It is to be noted that 
universal threshold corrections still leaves a possibility of
coupling constant unification within weakly coupled heterotic string
theory \cite{Nilles}.}. 

The above consideration points to the importance of a better 
understanding of M-theory compactification vacua. With such motivation, 
in the present paper, we study M-theory compactified on  Calabi-Yau 
three-folds $CY_3$  to five dimensions.
The compactification appropriately encodes the
strong coupling limit of type IIA compactification on the same 
Calabi-Yau three-fold  $CY_3$.  After the compactification,
the low-energy dynamics is governed by a $d=5, N=2$ supergravity. Its field
content and interactions are determined by the topological data of $CY_3$,
which may be understood, for example, from eleven-dimensional supergravity
compactified on $CY_3$\cite{Cadavid,FerraraKhuri}.

In understanding the web of moduli spaces of string compactification 
vacua, it has become clear that two aspects are essential. First, the 
classical moduli space may undergo topology changing phase transitions
accompanied by the appearance of massless states. Resolution of conifold
singularities~\cite{Strominger},
symmetry enhancement by small instantons~\cite{Witteninstanton}
and the 
appearance of tensionless strings~\cite{Wittendynamics,Ganortensionless,
Seibergdynamics,Dufftransitions,Reyprep} are well-known instances.
At the same time, it has become clearer that
BPS black holes with finite horizon serve as a spacetime probe of 
the compactified internal space~\cite{Ferraraextremal,Ferrarauniversality,
Kalloshfreezing,Behrndtstu,Behrndtclassical,Reyclassical,Behrndtstatic}. 
For example, for $d=4, N=2$ supergravity theory,
the so-called doubly extreme black holes attract K\"ahler
moduli scalar fields to a critical point of the central 
charge, $\partial Z / \partial  \phi^i = 0$.
The extremized BPS black hole mass then  becomes equal
 to the square root of the entropy, and depends only
on topological data of the 
compactified space such as intersection numbers, the second
Chern class and rational, elliptic instanton numbers.

In five dimensions, Witten~\cite{wittenM}
 has shown that all phases of the M-theory
compactification  vacua are purely geometrical. 
 Since the K\"ahler moduli do not admit
complexificiation in five dimensions, the
topology changing phase transitions in five dimensions are sharp. 
As such, the nature of phase transitions is quite distinct
from that in four dimensions. 
It is therefore of interest to 
understand all distinct features of the phase transition  explicitly
by means of simple probes. As in thermodynamical systems,
the nature of phase transitions and the corresponding
critical behavior can be probed by
various  supersymmetry potentials  of Landau-Ginzburg type. 
The BPS central charge $Z$, defined as the electric charge of the
gravi-photon,  which is the BPS mass  of the electric 5d black holes,
is one of such potentials. 
Similarly, the
magnetic charge of the gravi-photon $Z_m$  defines the tension 
of the magnetic 5d string and also can be used to probe the phase transition.
 Likewise, the black hole potential $V$ and the gauged central charge
potential $P$ are other additional physical quantities that can be used for 
analyzing the  topology-changing phase transitions. All these potentials
are defined for  fixed electric charges 
(wrapping numbers of M-branes on two-cycles of the Calabi-Yau three-fold)
and may be regarded as
independent thermodynamic potentials  in the canonical
ensemble\footnote{
Yet another possible  supersymmetry potential is the 
topological free energy \cite{topfreeenergy} 
in which all  BPS electric and magnetic charges are summed over. In
this paper, we will mainly consider canonical ensemble approach and do
not consider the topological free energy.}.
The critical point in the vector moduli space
is determined entirely by the chosen charge configuration for the potentials.
 The idea is that, 
by considering different charge configurations, the critical configuration
of the K\"ahler moduli for each of the above physical quantities 
may be located at different phases of the Calabi-Yau three-fold,
hence, probing the nature of phase transitions.

This paper is organized as follows. In section 2 we review the 
compactification of M-theory on a Calabi-Yau threefold. The resulting
low-energy approximation, $d=5, N=2$  supergravity coupled to
abelian vector 
supermultiplets, is based on the structure of {\it very special
geometry}~\cite{deWit}.
In section 3,
we derive the stabilization
equations of five-dimensional vector moduli scalars for electrically
charged particle states and magnetic string states. The central
charge $Z$ evaluated at these fixed scalar  field values turns out to
be a minimum as expected.
In section 4, we introduce two five-dimensional potentials of interest :--
the black hole potential $V$ and the scalar potential $P$  which  arises
 in gauged
supergravities. The central charge $Z$ and these two  potentials
are explicitly discussed
in a simple example. Section 5  constitutes the main results of our study.
After reviewing  briefly
some basic notions of the five-dimensional extended
K\"ahler cone, we probe the topology-changing phase transitions
using the supersymmetry potentials $Z$,  $Z_m$, $V$ and $P$.
Particular emphasis is put on  flop transitions
connecting two adjacent K\"ahler cones. We analyze how 
$Z$, $V$ and $P$ are affected by it.
 We find that the electric central charge and its first two derivatives
are in fact continuous along the flop transition. Since, as we 
will also show, any extremum of $Z$ has to be a minimum  one finds
uniqueness of the critical point of $Z$ in the entire extended
K\"ahler cone. Whereas the magnetic central charge (BPS tension)
and its first derivative are continuous, also implying uniqueness of the
stabilization point in the extended K\"ahler cone, 
$V$ on the other hand shows a kink  at the flop transition. 
Before we conclude,
our results are verified by an explicit example in section 
6, where we study one particular extended K\"ahler cone in detail. 
The behavior of the potentials  is graphically displayed. 
In addition we find new critical points for the potentials $V$ and $P$.
We observe that extremal transitions in five dimensions 
between Calabi-Yau spaces with different numbers of K\"ahler
moduli  often involve tensionless strings.

\section{Very Special Geometry and Calabi-Yau Manifolds}
We begin by recalling relevant aspects~\cite{Cadavid,Antoniadisfive} 
of M-theory compactification on
a Calabi-Yau three-fold $CY_3$ with Hodge numbers $h_{(1,1)}$, $h_{(2,1)}$ and
intersection numbers $C_{IJK} (I,J,K = 1, \cdots, h_{(1,1)})$.
The bosonic fields of the supergravity multiplet in eleven dimensions 
consist of the metric $g_{MN}$ and the 3-form
antisymmetric tensor potential ${\cal A_{MNP}}$. 
Denote the complex coordinates of
$CY_3$ as ${\cal I}, {\cal J} , {\overline {\cal I}}, 
{\overline {\cal J}} = 1,2,3$.
After  compactification on the $CY_3$, the action contains one 
CY complex axion
${\cal A}_{{\cal I J K}} = \epsilon_{\cal I J K} b$, one universal 
axion $a \,\, (d a \sim {}^* d {\cal A})$,
the Calabi-Yau 
volume ${\rm det}~ {\cal G_{I {\overline J}}}$, $h_{(1,1)} - 1 $ real K\"ahler
moduli  ${\cal G_{I {\overline J}}}$ and $h_{(2,1)}$ pairs of complex
scalars $({\cal G_{IJ}}, {\cal A_{IJ {\overline K}}})$.
In addition, $h_{(1,1)}$ abelian vector potentials 
${\cal A}_{\mu {\cal I {\overline J}}}$
arise, one of which becomes the gravi-photon.
They combine into $d=5$ supermultiplets as follows.
Taking into account one graviphoton in the supergravity multiplet,
there are $h_{(1,1)}-1$ vector multiplets consisting of uni-modular
$({\cal G_{I {\overline J}}}, {\cal A}_{\mu {\cal I {\overline J}}})$, 
one universal hypermultiplet
$({\rm det}~ {\cal G_{I {\overline J}}}, a, 
{\cal A_{IJK}})$, and $h_{(2,1)}$ hypermultiplets
containing
$({\cal G_{ IJ}}, {\cal A_{IJ {\overline K}}})$.
Supersymmetry non-renormalization theorems~\cite{nonrenor, Becker} 
ensure that the moduli space of the
hypermultiplets containing the complex
structure moduli (and the size of the entire Calabi-Yau space) 
is completely disconnected from the K\"ahler moduli space.

In this paper we will restrict ourselves to the moduli space of
the vector multiplets, corresponding to the K\"ahler moduli $t^I$ of the
Calabi-Yau space. The K\"ahler moduli are derived from the
K\"ahler form $J$ on $CY_3$  by expanding into a fundamental  2-form
basis $J_I$ as
\be
J=t^I J_I.
\ee
With this definition the moduli $t^I$  correspond to the
sizes of 2-cycles in $CY_3$ dual to the $J_I$.

The supersymmetric coupling of $d=5, N=2$ vector multiplets to N=2
supergravity  has been constructed 
in
\cite{Gunaydingeometry}\footnote{ We adopt different normalization of
the action  from the earlier works 
\cite{Antoniadisfive,Chamseddinefive}
so that the Chern-Simons coupling $C_{IJK}$ equals directly to
the topological intersection number on Calabi-Yau three-folds.}.
The vector multiplet coupling is determined completely by specifying a
prepotential
\begin{equation}
{\cal V} = C_{IJK} t^I t^J t^K,
\hskip1cm I = (1, \cdots, h_{(1,1)}),
\label{prep}\end{equation}
which defines intrinsic `very special geometry' \cite{deWit}.
The prepotential is related to the overall volume of Calabi-Yau three-fold.
In five dimensions, the overall volume  belongs to the 
so-called universal
hypermultiplet. Therefore, the prepotential which encodes independent
degrees of freedom of the vector multiplet moduli space is constrained to
satisfy
\be
\Nu =1.
\ee
The variables $t^I$ are related to the conventional
very special coordinates \cite{Chamseddinefive} $X^I$ by the rescaling
\be
t^I=6^{\frac{1}{3}}X^I.
\ee
The $h_{(1,1)}-1$ independent, real scalar fields parametrizing
the vector multiplet moduli space are denoted by $\phi^i$, and the
$t^I$ are functions of them. It is often convenient 
to identify the $\phi^i$ with $h_{(1,1)}-1$
of the $t^I$ and to take the remaining K\"ahler moduli to be
a function of the other ones defined by the constraint $\Nu=1$.

In terms of the aforementioned normalization convention,
the bosonic part of the action is given by
\begin{eqnarray}
e^{-1} {\cal L} = -{1\over 2} R   - {1\over 4} G_{IJ}(\phi)
F_{\mu\nu} {}^I
F^{\mu\nu J}
 -{1\over 2} g_{ij}(\phi) \partial_{\mu} \phi^i \partial^\mu \phi^j
+{ e^{-1}  \over 48} \epsilon^{\mu\nu\rho\sigma\lambda} C_{IJK}
F_{\mu\nu}^I
F_{\rho\sigma}^J A_\lambda^K  .
\label{veryspecial}\end{eqnarray}
The vector multiplet coupling is defined by
\be
G_{IJ}=-\frac{1}{2\cdot 6^{\frac{2}{3}}} 
   \partial_I \partial_J (\ln \Nu) |_{\Nu=1},
\ee
while the moduli space metric $g_{ij}$ is given by
\be
g_{ij}=6^{\frac{2}{3}}
   G_{IJ} t^I_{,i}t^J_{,j}=-3 C_{IJK}t^I t^J_{,i} t^K_{,j}.
\ee
The  scalar fields $\phi^i, \,\, (i=1, \cdots, h_{(1,1)}-1)$ parametrize
the vector multiplet moduli space
${\cal M}_V$. For a given ${\cal M}_V$, consider the second covariant derivative of
the very special coordinate with respect to the moduli scalar fields $\phi^i$
\begin{equation}
(t^I)_{,i;j} =  (2/3) g_{ij} t^I -  \sqrt { 2/3} \;T_{ijk } t^{I , k}.
\label{second}\end{equation}
The symmetric tensor $T_{ijk}$, which is a function of the moduli scalar
fields, encodes local geometric structures of ${\cal M}_V$\footnote{
The same tensor
appears in a number of  places in the Lagrangian including 4-fermion terms.}.
If $T_{ijk}$ is covariantly constant, $T_{ijk;l}=0$, the scalar
manifold ${\cal M}_V$ is locally a symmetric space. $d=5, N=2$ supergravity
with locally symmetric ${\cal M}_V$
has been studied extensively with the aid of Jordan algebras
\cite{Gunaydingeometry}. Generically, with $n \equiv h_{(1,1)}-1$,
\begin{equation}
{\cal M}= {SO(1,1)} \times {SO(1, n-1)\over SO(n-1)}.
\end{equation}
In this case the prepotential ${\cal V} = C_{IJK} t^I t^J t^K $ should be
factorizable into a linear and a quadratic polynomial parts. That is,
\begin{equation}
{\cal V} = C_{IJK} t^I t^J t^K = t^1 Q(t^\lambda)
\label{fact}\end{equation}
 where $\lambda=2,..,n $ and $Q$ is a quadratic form of
signature $(+,-,-,\dots, -)$.
If ${\cal V}$ is not factorizable, ${\cal M}_V$ may or may not
be a symmetric space. The cases with
locally symmetric and homogeneous manifold ${\cal M}_V$, but with
non-factorizable prepotential ${\cal V}$ have been classified in
\cite{Gunaydingeometry}. There are only four exceptional cases,
 so-called `magic square' supergravities, for which
$n= 3(1+ {\rm dim} {\bf A})$ where ${\bf A}$ is one of the division
algebras and $n=5, 8, 14, 26$.

In generic Calabi-Yau compactifications of M-theory, the
prepotentials do not belong to the set based on
the Jordan algebra classification of $d=5$
supergravities. Possible exceptions are Calabi-Yau compactifications
described by factorizable prepotentials of the form (\ref{fact}) or
 the `magic square' cases $n=5,8,14, 26$.
In all other cases, the $d=5$ compactifications
do not share
many of the properties of the known $d=5$ supergravities
based on  the Jordan algebras.
In following sections, we will study Calabi-Yau prepotentials
with small dimensions of vector moduli spaces
$n = 1, 2$,
which are neither factorizable nor belong to the ``magic square" cases.
As we will see, these examples are simple enough to allow a detailed analysis,
while rich enough to capture all the essentials of  geometric
phase transitions  in five dimensions.

\section{ Stabilization Conditions }
In $d=5, N=2$ supergravity theories, the gravi-photon field $T$
is given by a linear combination of all vector fields $F^I$ of the
theory with  moduli dependent coefficients: $T= t^I  (\phi)
G_{IJ} (\phi)
F^I$. Associated to the gravi-photon is the electric central charge
\cite{Antoniadisfive}
\begin{equation}
Z=  t^I (\phi) q_I
\end{equation}
where $q_I$ is the quantized electric charge for each vector field
$G_I \equiv G_{IJ} F^I$. In the context of M-theory the electrically
charged point-particle states are identified with
BPS winding states of the  M-theory
membrane around 2-cycles  of the
Calabi-Yau three-fold $CY_3$ \cite{Cadavid,Antoniadisfive,
Becker}. Magnetically charged string states are
identified with BPS winding states of the  M-theory
five-brane around
4-cycles. The magnetic central charge  describing the 
string tension is given by \cite{Antoniadisfive}
\begin{equation}
Z_m=  t_I (\phi) m^I
\ee
where the dual moduli $t_I$ defined by
\be
t_I=C_{IJK}t^J t^K
\la{tdual}
\end{equation}
encode the sizes of four-cycles  of the $CY_3$.
The mass spectrum of BPS states varies over the vector multiplet moduli space.
As such, the mass of a BPS state can often be extremized as a function of
moduli for given values of charges $q_I$ \cite{Ferrarauniversality,
Ferraraattractors}.
It is therefore of interest to study critical behavior of
the BPS mass-squared over the moduli space ${\cal M}_V$
\begin{equation}
M^2_{\rm BPS} = Z^2 (q_I, C_{IJK}, \phi^i) = t^I t^J q_I q_J.
\end{equation}
The critical points of the BPS mass-squared are specified by the extremal
value of the central charge $\partial_i Z=0$
or by the vanishing  of the central charge  $ Z=0$
since
\begin{equation}
{\partial M^2_{\rm BPS}\over \partial \phi^i}   = 2 Z \, \partial_i Z = 2
(t^I)_{,i} \; q_I  \; t^J  q_J =0.
\end{equation}
In the $d=4, N=2$ theory the extremal configuration of the central charge
$\partial_i Z=0$ has been
found to be a solution of a set of so-called stabilization equations,
which define  functional relations between microscopic charges
and moduli scalar fields at the critical points
\cite{Ferrarauniversality,Ferraraattractors}. Quite often, the stabilization
condition equations are easier to solve than the problem
of extremizing directly the central
charge \cite{Kalloshfreezing,Behrndtstu,Behrndtclassical,Reyclassical,
Behrndtstatic}.

In what follows, we derive the corresponding stabilization condition equations
for $d=5, N=2$ supergravity in two alternative ways.
The first derivation extends the method utilized in the $d=4, N=2$ context
\cite{Ferrarauniversality,Ferraraattractors}:
the very special geometry that underlies the vector moduli space is
used to invert the  implicit functional relations
 $\partial_i Z=(t^I) _{,i} q_I = 0$ into equations
between charges and moduli. The second method employs a Lagrange multiplier
to the  constraint equation ${\cal V}(t)  =  C_{IJK} t^I t^J t^K =1 $
and also
effectively inverts the $(t^I) _{,i}q_I =0$ condition.

\subsection{Stabilization of the Electric Central Charge:
Very Special Geometry}
To invert  $\partial_i Z=(t^I) _{,i}q_I =0$ and find the expression for $q_I$'s
in terms of $t^I$'s we will utilize some of the properties of the
very special geometry presented in \cite{Chamseddinefive}.
We first multiply $\partial_i Z $ by $g^{ij}  t^J{}_{,j}$ and get
\begin{equation}
   \partial_i Z=(t^I) _{,i}q_I =0   \qquad  \Longrightarrow  \qquad  g^{ij}
t^I{}_{,i} t^J{}_{,j} \,q_I
\equiv \Pi ^{IJ} q_I  = -{1\over 3} ((C^{IJ}  -  t^I t^J)q_I =0 
\label{identity1}\end{equation}
where the matrix $C^{IJ}$ is the inverse matrix of $C_{IJ} = C_{IJK} t^K$.
Multiplying this equation by $C_{KJ}$, we find that
\begin{equation}
q_K = C_ {KJ} t^J  Z = C_ {IJK} t^J t^K  Z
\end{equation}
at $\partial_i Z = 0$.
Thus the $d=5$ stabilization condition equations have a particularly simple
form
 \begin{equation}
q_I  = Z  t_I 
\end{equation}
where $t_I$, as defined in (\ref{tdual}),
is the coordinate dual to the very special coordinate $t^I$.

\subsection{Alternative Derivation: Lagrange Multiplier Method}

We use the linear sigma model approach 
to the constraint of the prepotential.
Consider a function:
\begin{equation}
\tilde Z (t^I) = Z +\lambda ( {\cal V}-1) = q_It^I+\lambda
(C_{IJK}t^It^Jt^K
-1),
\end{equation}
where the $t^I$'s are {\it a priori} 
unconstrained and $\lambda$ denotes a Lagrange
multiplier.
Taking derivatives with respect to $t^I$ we obtain a set of extremization
equations:
\begin{equation}
q_I+3 \lambda C_{IJK}t^Jt^K = 0. \label{Lagrange}
\end{equation}
Similarly, the extremization condition with respect to $\lambda$ gives:
\begin{equation}
C_{IJK} t^I t^J t^K = 1. \la{constraint1}
\end{equation}
Contracting (\ref{Lagrange})
with $t^I$ and using the constraint (\ref{constraint1})
we  solve for
$\lambda$ as
\begin{equation}
3 \lambda = - q_I t^I = -Z.
\end{equation}
Substituting (\ref{constraint1}) into (\ref{Lagrange}) we obtain
\begin{equation}
q_I-Z C_{IJK}t^Jt^K=0.
\end{equation}
In terms of the dual variable $t_I$,  the solution is
\begin{equation}
q_I= Z t_I,
\end{equation}
yielding the same result as the very special geometry method.

For later use, we find it useful to put the $d=5$ stabilization equation
with a suitable rescaling. Let
us call $\bar t^I$ the original very special coordinate rescaled
 by the
square root of the central charge
\begin{equation}
\bar t^I \equiv  \sqrt {Z } \; t^I.
\end{equation}
Since the coordinates $t^I$ of the very special geometry  are real,
the rescaled coordinates $\bar t^I$ have to be  real  for
configurations with positive value of the central charge.
In $d=5$, one can also consider the case of imaginary  $ \bar t$ with
negative
$Z$. One can see however that this case is
easily reproduced starting with the configuration with positive $Z$
and real solutions for   $ \bar t$ by  flipping the signs of all
charges, as the critical points of $Z$ and $-Z$ do coincide manifestly.
It is then possible to rewrite the $d=5$ stabilization equation into a rather
suggestive form
\begin{equation}
q_I=  C_{IJK} \bar t^J  \bar t^K.
\label{stab}\end{equation}
This equation implies that, for a given choice of Calabi-Yau manifold with
specific topological intersection numbers and a given number of moduli
fields, the value of
the electric charge $q_I$ defines the real (imaginary) values of rescaled
moduli  via eq.(\ref{stab}),  
which is quadratic in the rescaled moduli. The
solution may or may not exist for the real (or purely
imaginary) $\bar t^I$. In cases where the solution exists we
have an explicit expression for the
extremum of the BPS mass and for the black hole entropy. Once the
solution to
eq. (\ref{stab}) is found in the form $\bar t^I (q_L, C_{IJK}) $
we find
the  critical value of the central charge in the form
\begin{equation}
\left(\bar t^I q_I \right)^2 =  Z^{3}_{\rm cr} (q_L, C_{IJK}).
\label{BPS}\end{equation}
This in turn defines the entropy of the black hole
\cite{Ferrarauniversality,Ferraraattractors,Chamseddinefive}
\begin{equation}
S  (q_L, C_{IJK})  = {\pi^2 \over 12} |Z_{\rm cr}|^{3/2}.
\label{entropy}\end{equation}
Recall that the near horizon geometry of the $d=5$ black hole is
given by $adS_2 \times S^3$, hence, the black hole entropy is
proportional to the volume of the $S^3$.
The expression for the critical value of the central charge has been
found previously
\cite{Ferrarauniversality,Ferraraattractors,Chamseddinefive}
in the form $ (Z)^{2}_{\rm cr} = q_I q_J (C^{IJ})_{\rm
cr}$. The
new $d=5$ stabilization equation (\ref{stab}) enables
to find the entropy of $d=5$ black holes explicitly as a function of
the microscopic electric charges and of the CY topological intersection
numbers.

Indeed, the structure of the $d=5$ stabilization equations
are far simpler to solve than those in the $d=4$ case. In \cite{Shmakova}
it was  found that the  $d=4, N=2$ black hole entropy with classical
Calabi-Yau prepotential can be understood completely once
 a real solution is found to the following set of
algebraic equations:  $\Delta_I=  C_{IJK} \tilde x^J  \tilde x^K
$ where
$\Delta_I$ is some particular combination of microscopic electric and
magnetic charges and topological intersection numbers. When these equations
have a solution with real $ \tilde x^J$, an algorithm was presented for
finding the fixed values of moduli,
the central charge and
the entropy of black holes. The algorithm was general enough but
the explicit form of the result was rather complicated.

In $d=5$ the situation has been simplified considerably.
The
stabilization
equations $q_I=  C_{IJK} \bar t^J  \bar t^K$  coincide with
the algebraic equations of \cite{Shmakova} $\Delta_I=  C_{IJK} \tilde x^J
\tilde x^K
$ upon
identification $\Delta_I \sim q_I$ and $ \bar t^J  \sim  \tilde
x^J$.
   Moreover, the algorithm for obtaining the
critical value of
the BPS mass   and the entropy after $\bar t$  is found is
very simple in
$d=5$ and displayed in eqs. (\ref{BPS} ) and  (\ref{entropy}).
If one would like to use the critical value of the $d=4$ central
charge directly
to
find the critical value of its 5d counterpart the following
restrictions have to be applied to the  $d=4$ 
charges:  $p^I= q_0= p^0-1 =0$.

\subsection{Stabilization of the Magnetic String Tension}
The magnetic central charge determining the tension of the magnetic 
string states
\cite{Antoniadisfive} is a function of the moduli as well, hence,
one can also have a critical point for fixed values of the microscopic
magnetic charges $m^I$
\begin{equation}
\partial_i  Z_m = (t_I)_{,i} m^I = 2 C_{IJK} t^I (t^J)_{,i} m^K =0.
\end{equation}
To invert this relation we proceed similarly as before
\begin{equation}
 C_{IJK} t^I (t^J)_{,i}  (t^L)_{,j}  g^{ij} m^K = C_{IJK} t^I m^K
(C^{JL} - t^J
t^L) =0.
\end{equation}
It follows that
\begin{equation}
m^L = t^L Z_m  \qquad \Rightarrow \qquad t^L = {m^L \over Z_m}.
\end{equation}
The critical value of the dual coordinate is
\begin{equation}
t_I  = {C_{IJK} m^J m^K  \over Z^2_m}
\end{equation}
and the critical value of the BPS string tension is
\begin{equation}
 (Z_m ) ^3 _{\rm cr} = (t_L m^L )^3 _{\rm cr} = C_{IJK} m^I m^J m^K.
\end{equation}
For the electrically charged $d=5$ BPS black holes, the critical value of
the BPS mass was related to the horizon volume $S^3$, where the near
horizon geometry was given by $adS_{2} \times S^3$.
This in turn was proportional to the black hole entropy.
For the magnetically charged $d=5$ BPS black string,
we  find similarly that the extremized value of the BPS
tension is related to the volume of the $S^2$ as well as to the size 
of the infinite throat  of   $adS_{3}$ where the near horizon geometry is
given by
$adS_{3} \times S^2$. The volume of $S^2$ is moduli independent
and is proportional to $Z_m ^2 = ( C_{IJK} m^I m^J m^K)^{2\over 3}$.
For  $d=4$ black holes
 the relevant no-axion black hole solution with the entropy
proportional
to $\sqrt {q_0 C_{IJK} m^I m^J m^K }$  was found  in \cite{Behrndtclassical}
and  an empirical rule of state counting in M-theory was
proposed in
\cite{Behrndtentropy}, \cite{Maldacenaentropy} to explain the entropy.
Upon decompactification  to $d=5$ this black hole solution was
 identified with
the magnetic string in \cite{Behrndtdeco}. Here we have found that the
stabilization
of the magnetic string tension in $d=5$ describes the same structure as the
stabilization of the magnetic no-axion black hole in $d=4$ with unit
electric
charge $q_0=1$. 
The critical value of the 5d magnetic tension
can be obtained from the  stabilization of the 4d black hole 
mass at $q_0-1 = q_I = p^0 =0$. This result
follows simply from comparing $d=4$ results for the black hole with the 
$d=5$ results for the magnetic string.

\subsection {The Extremum of the BPS Mass is a Minimum}
A natural question to address is under what conditions the extremum
of the BPS mass is a {\sl minimum}. An analogous issue  has been
already addressed in
the  context of $d=4$ special geometry \cite{Ferraracritical},
where it was found that
\be
D_i D_j Z^2 =\frac{1}{2} g_{ij} Z^2 \qquad {\rm at} \qquad \partial_i Z=0.
\ee
In $d=5$ the
corresponding calculation yields, using eq. (\ref{second}):
\begin{equation}
D_i D_j Z = \partial_i \partial _j Z= {2\over 3} g_{ij}
Z_{\rm cr}
\qquad {\rm at}  \quad \partial_i Z =0 \ .
\end{equation}
This shows that the extremum of the central charge is a minimum (maximum)
if the central charge at the critical point is positive (negative),  
using the fact  that the metric of the moduli space is positive definite.
For the square of the BPS mass we get
\begin{equation}
D_i D_j (Z)^2 = \partial_i \partial _j (Z)^2 = {4\over 3} g_{ij}
(Z^2)_{\rm cr}
\qquad {\rm at}  \quad \partial_i Z =0 \ .
\end{equation}
The same is true for the magnetic BPS mass.
Thus, as long as the very special geometry is regular  viz.  $g_{ij}$
and $G_{IJ}$
are positive
definite and non-degenerate, the BPS mass has a minimum at the
critical point.
Hence, for a given choice of charges and signs of moduli at infinity 
the corresponding minimum, if it exists,  is unique.  

\section{The Black Hole and Gauged Central Charge Potentials}
In the last section we  have studied the critical
points of the central charge $Z$  in detail. For
black holes, a quantity of more direct significance  is
the black hole potential $V$. In this section we introduce
this potential  along with a  gauged central charge
potential $P$, which arises
in gauged supergravities, and discuss their respective critical
points in detail. We will also compare the 
critical behavior of $Z^2$, $V$ and $P$
for an example defined by prepotential $\Nu=STU$.
  
\subsection{The Black Hole Potential and Its Critical Points}
In the context of $d=4,  N=2$ supergravity, it was explained in
\cite{Ferraracritical} that the stabilization of moduli fields of
BPS black holes near
the black hole horizon
depends on the properties of the so-called black hole potential $V$. 
For extremal (but not necessarily supersymmetric) 
black holes with regular horizon and regular moduli scalars,   
the following
relations have been established.  The entropy is defined (in eq. (26) of
\cite{Ferraracritical})
as the value of the black hole potential evaluated at the
horizon:
\begin{equation}
{S\over  \pi} = V(p,q, \phi_h).
\end{equation}
Furthermore, in  \cite{Breitenlohner,Ferraracritical,Gibbonsmoduli},
it was found that the ADM mass of
a generic non-extreme black hole consists of  three
 contributions:  the
black hole
potential depending on the value of moduli at infinity, some
function of the
scalar charges and the $(2ST)^2$ term, where $T$ is the black hole
temperature:
\begin{equation}
(M_{\rm ADM})^2  =  V(p,q, \phi_\infty) - G_{ab} \Sigma^a \Sigma^b
+ (2ST)^2.
\end{equation}
For the case of $d=4, N=2$ 
supergravity, utilizing  special geometry relations,
$V$ was found
to be equal to $|Z|^2+ |\partial Z|^2$.  In the extreme case when
$TS=0$ there are two possibilities: (1) the extreme black hole is
supersymmetric
\begin{equation}
(M_{\rm ADM})^2 = (M_{\rm BPS })^2 = |Z(p,q, \phi_\infty)|^2,
\end{equation}
which requires for consistency that
$
|\partial Z(p,q, \phi_\infty)|^2 = G_{ab} \Sigma^a \Sigma^b
$
and the entropy is given by ${S\over  \pi} = V(p,q, \phi_h)= |Z|^2
$ at $\partial_i |Z|=0$, or
(2) the extreme black hole is not supersymmetric
\begin{equation}
(M_{\rm ADM})^2  =  |Z|^2+ |\partial Z|^2 - G_{ab} \Sigma^a
\Sigma^b >  |Z(p,q,
\phi_\infty)|^2 = (M_{\rm BPS })^2
\end{equation}
and the black hole horizon area is still given by the value of the
potential at the horizon ${S\over  \pi} = V(p,q,
\phi_h)$ if the moduli are regular (in 
such cases the critical point of the black hole
potential is not defined by the
enhancement of unbroken supersymmetry ($\partial_i |Z|=0$) as it
happens for
supersymmetric black holes). The potential has a critical point
which describes
the stabilization of moduli near the horizon of extreme
non-supersymmetric
black holes. These non-supersymmetric critical points have not been studied
much previously. However, the extreme
non-supersymmetric solutions with (or without) regular horizon
have been known to exist \cite{Giddings,Khuri,Duffbound,Duffgyro}.
The most recent detailed study of such solutions in
$N=4$ and $N=8$ supergravity theories  was given in \cite{Ortin}.

In what follows we will study the analogous black hole potential in
$d=5$.  We
will find as expected both types of critical points. The
gauge  kinetic term
in the Lagrangian responsible for the black hole potential is
proportional to
$G_{IJ} F^I F^J$.
Using  eqs. (16,29) from \cite{Chamseddinefive} 
one finds that the $d=5$ potential whose
critical points describe the BPS mass and the stabilization of
moduli for
extreme black holes is given by
\begin{equation}
V= Z^2 +  {3\over 2}  g^{ij} \partial_i Z \partial_j Z = \left (t^I t^J + {3\over 2} g^{ij}
(t^I)_{,i} ( t^J)_{,j}\right) q_I q_J.
\end{equation}
 It can be shown that the potential is 
proportional to the gauge couplings
\begin{equation}
V = {6^{1/3}\over 4} G^{IJ}(\phi)  q_I q_J
\end{equation}
and the critical points of this potential are given by
\begin{equation}
\partial _i V = 4 \partial _i Z Z  -  \sqrt {6} \;  T_{ikj}
\partial ^k Z
\partial ^j Z=0.
\end{equation}
There can be

1. {\bf Supersymmetric critical points}.
Obviously the critical point of the central charge is also a
critical point
of the potential
\begin{equation}
\partial _i V = 0  \qquad {\rm at} \quad  \partial _i  Z =0.
\label{susy}\end{equation}
 This applies to  supersymmetric extreme black holes.
The inverse is not true: it is possible that there is a

2. {\bf Non-supersymmetric critical point} where
\begin{equation}
\partial _i V  = 0  \qquad {\rm at} \quad    \partial _i  Z  \neq 0
\label{nonsusy}\end{equation}
which give rise to stabilization of the moduli also for non-supersymmetric
black holes.
The calculation of the second derivative of the potential at the
supersymmetric critical point is straightforward and is given  by
\begin{equation}
(V) _{,i; j}= \partial_i  \partial_i V = {8\over 3} g_{ij} V_{\rm
cr} \qquad
{\rm at} \qquad \partial_i V = \partial_i Z  =0.
\end{equation}
We conclude that for  a positive definite metric of  very special geometry
the potential has a minimum at the supersymmetric critical point (\ref{susy}), 
which exists under the condition that  the stabilization
equation  (\ref{stab}) for the moduli in terms of charges has a real
solution.

The non-supersymmetric critical points (\ref{nonsusy}) do not seem
to allow a simple
general treatment. We will study them  for a few  
examples and show that indeed
such  critical points do exist.

\subsection{The Gauged Central Charge Potential}
The gauged central charge of $N=2$ supergravity arises in theories
in which an $U(1)$ subgroup of the $SU(2)$ automorphism group of the $N=2$
supersymmetry algebra is gauged \cite{Gunaydingeometry}.
The gauging which breaks $SU(2)$ down
to $U(1)$ is realized by  making the  gravitino and gaugino 
charged
so that their covariant derivatives are
\begin{eqnarray}
(D_\mu \lambda^i)^\alpha &=& \nabla _\mu \lambda^{i^\alpha} + g V_I
A_\mu {}^I
\delta^{\alpha \beta} \lambda^i{}_{\beta} \\
(D_\mu \psi_\nu )^\alpha &=& \nabla _\mu \psi_\nu ^{\alpha} + g V_I
A_\mu {}^I
\delta^{\alpha \beta} \psi_\nu {}_{\beta}\ ,  \qquad \alpha, \beta =1,2.
\end{eqnarray}
To maintain the $N=2$ supersymmetry intact after the gauging,
one has to add to the theory  the gauged central charge
 potential proportional to  $g^2 P$, where
\begin{equation}
P=  Z^2- {3\over 4}  g^{ij} \partial_i Z \partial_j Z.
\end{equation}
In the context of gauged supergravity the central charge is
actually a moduli
dependent combination of gravitino and gaugino charges defined by
\begin{equation}
Z = t^I V_I
\end{equation}
where $V_I$ is the charge defining the gravitino-gravitino-vector and
gaugino-gaugino-vector interactions as may be deduced from their
respective covariant
derivatives.
In addition to the potential, there are gravitino and gaugino mass
terms proportional to
\begin{eqnarray}
L_{\rm mass}^{\rm grav} &=& - i \bar \psi  ^\alpha_{\mu}
\Gamma^{\mu\nu}   \psi
 ^\beta_{\nu} \delta_{\alpha \beta} Z\\
L_{\rm mass}^{\rm gauge} &=& - i \bar \lambda  ^{\alpha i}
\lambda  ^{\beta
j} \delta_{\alpha \beta} \left( g_{ij} Z + 4\sqrt 2 T_{ijk} Z^{,k}
\right)
\end{eqnarray}
and a mixing term proportional to the derivative of the central charge
$\bar \lambda  ^{\alpha i} \Gamma^{\nu}   \psi  ^\beta_{\nu}
\delta_{\alpha
\beta}  Z_{,i}$.
Thus in this theory the masses of gravitino and gaugino coincide
with  the BPS
mass in the generic point of the moduli space up to terms
proportional to the
derivative of the central charge in the moduli space. This observation
implies that  at
supersymmetric
critical points where $Z_{,i}=0$, one finds
\begin{equation}
M^{\rm grav}  = M^{\rm gauge} =|(t^I )_{\rm cr} V_I |=  |Z_{\rm cr}  (V_I,
C_{IJK})| \qquad  {\rm at} \qquad  Z_{,i}=0.
\end{equation}
It is interesting to note that, as a function of gravitino
charges and  topological
 intersection numbers, the gravitino  mass 
is given by the same
topological formula which defines the entropy of the extreme
supersymmetric
black holes as function of black hole charges.
The critical points of the $P$ are given by
\begin{equation}
\partial _i P =  Z \, \partial _i Z   +  \sqrt {3/ 2} \;  T_{ijk}
\partial ^j Z
\partial ^k Z=0.
\end{equation}
As in previous cases it is straightforward to study
supersymmetric critical
point where the second derivative of the potential is proportional
to the
metric on the moduli space:
\begin{equation}
(P) _{,i; j}= \partial_i  \partial_i P = {2\over 3} g_{ij} (Z^2)_{\rm cr}
\qquad {\rm at} \qquad \partial_i P= \partial_i Z  =0.
\label{critP} \end{equation}
Within the validity range of very special geometry,
this potential has a minimum
whenever the stabilization equation has a solution. 
In the case of  potentials related to 
Jordan algebras all critical points of  $P$
 have
been classified in \cite{Gunaydingeometry}, where two types of
critical points were described.
It was found that the first derivative of the potential  vanishes
when the
configuration of charges is such that either the derivative of the
central
charge is vanishing  in our formalism or the potential is vanishing
\begin{equation}
\partial _i P \sim   \partial _i Z\: P =0 \qquad {\rm at} \qquad
\partial_i Z
=0 \quad {\rm or} \qquad P  =0.
\end{equation}
In the first case, the potential is non-zero at the minimum
giving rise to a cosmological constant. This critical point
is  associated with 
supersymmetric black
holes and extremization of their mass. The second case with the
identically
vanishing  potential (cosmological constant) is specific to the theories
associated with irreducible idempotents of the Jordan algebra.

As explained in section 2, M-theory compactified on a $CY_3$ manifold
typically does not
give rise to a factorizable prepotential nor corresponds to
one of the magic square supergravity theories.
Properties of the supersymmetric critical points of $V$ and $P$ can be studied
using eqs. (\ref{susy},\ref{critP}). However, critical points of any 
other nature
are neither known nor excluded for the two  potentials in case generic
 $CY_3$ compactifications.
We will see however in section 6 by analyzing specific
examples that generically there exist additional critical
points.

To become familiar with the various kinds of potentials  discussed
so far
we will examine a simple example defined by  $\Nu=STU$.
Although this prepotential does not directly 
correspond to a Calabi-Yau compactification it 
is useful enough to illustrate some  essential ideas.

\subsection{The ${\cal V}=STU=1$ Example}
This example corresponds to a factorizable case, hence, the
full Jordan algebra apparatus applies and in fact this prepotential
 has been studied in detail in
\cite{Gunaydingeometry}. The solutions to the stabilization
equations are given by
\bea
S&=& (\frac{q_T q_U}{q_S^2})^{\frac{1}{3}} \nn \\
T&=& (\frac{q_U q_S}{q_T^2})^{\frac{1}{3}}  \la{stabstu}\\
U&=& (\frac{q_S q_T}{q_U^2})^{\frac{1}{3}}  \nn
\eea
and lead to the central charge
\be
Z=3(q_S q_T q_U)^{\frac{1}{3}}.
\ee
At this point we have not explicitly required positivity
of the moduli fields.  The result  (\ref{stabstu}) shows that
we find exactly one critical point of the central charge
for a given set
of charges. However, only if all of them are positive
corresponding to no anti-brane configurations,
we find the stabilization of the central charge for positive values
of the moduli fields.

To understand the structure of the moduli space it is also
instructive to analyze the moduli space metric $g_{ij}$ and the gauge
couplings $G_{IJ}$. Taking $S$ as the dependent field, we find
\bea
g_{(TU)}&=&\pmatrix{\frac{1}{T^2} & \frac{1}{2 TU} \cr
\frac{1}{2 TU}& \frac{1}{U^2} \cr
} \nn \\
G&=&\frac{1}{2\times 6^{\frac{2}{3}}}\pmatrix{\frac{1}{T^2} & 0 & 0 \cr
0 &\frac{1}{U^2}  & 0 \cr
 0 & 0 & T^2 U^2 \cr}.
\eea
Clearly, whenever either of the boundaries $T=0$ or $U=0$ is approached
the corresponding  matrix elements in $g$ and $G$ diverge. 
This implies that
the boundaries are infinitely far away from a generic point in the
moduli space.  Also, it
 is to be noted that the vector field $F_S$ becomes
strongly coupled  at the boundaries.

The potentials are easily calculated. We find
\bea
Z^2&=&(\frac{q_S}{TU}+ q_T T +q_U U)^2 \nn \\
V&=&3\left((\frac{q_S}{TU})^2+ (q_T T)^2 +(q_U U)^2
\right)\la{stupots} \\
P&=& 3( T U q_T q_U + \frac{q_S q_U}{T} + \frac{q_S q_T }{U}). \nn
\eea
As discussed earlier the stabilization equations provide  only
critical points of $Z$ (and $Z^2$) which in turn are also critical
points of $V$ and $P$. However, already in this example we find new
stabilization points of $V$. To see this
consider for example the case of $q_T,q_U<0, q_S>0$. The result of
(\ref{stabstu}) indicates the existence of an extremum only in the
$U,T<0$ domain. On the other hand, since in $V$ the charges  arise
in quadratic form only, we do find an extremum also in the
positive domain of $U$ and $T$ as well. Hence, critical points of
the potential can also occur when $\partial_i Z\neq 0$.
This behavior is visualized in Figure 1.
We plot contour graphs for $Z^2,V$ and $P$ with charge configurations
$(q_S,q_T,q_U)=(1,1,1),(1,-1,1),(1,-1,-1)$ in the neighborhood
of $(T,U)=(1,1)$.
Indeed, the central charge and $P$   stabilize
for positive charges only,  whereas $V$ has  always a minimum.
\begin{figure}
$$
\begin{array}{cccc}
Z^2 &  
\epsfxsize=30mm\epsfbox[0 100 200 200]{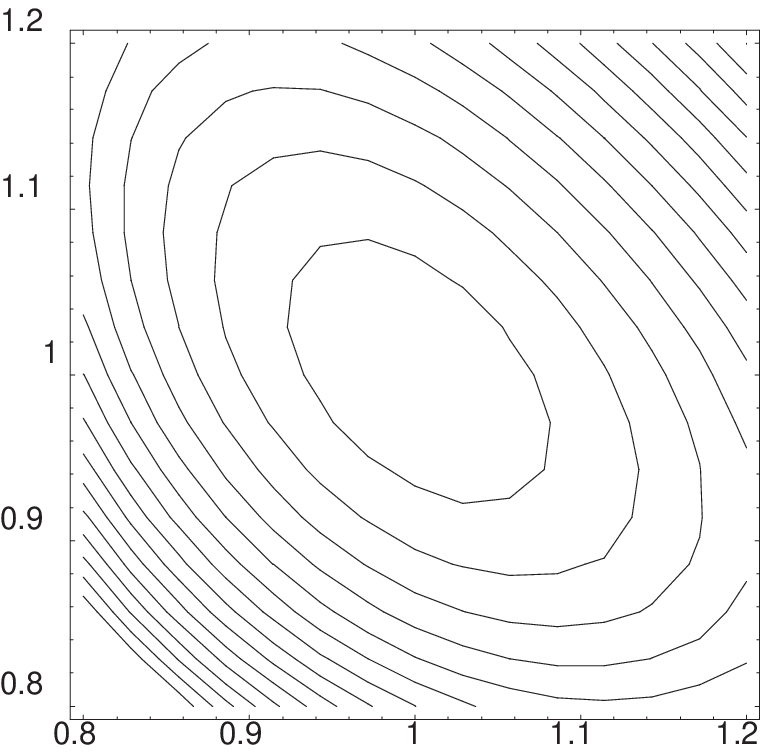} & 
\epsfxsize=30mm\epsfbox[0 100 200 200]{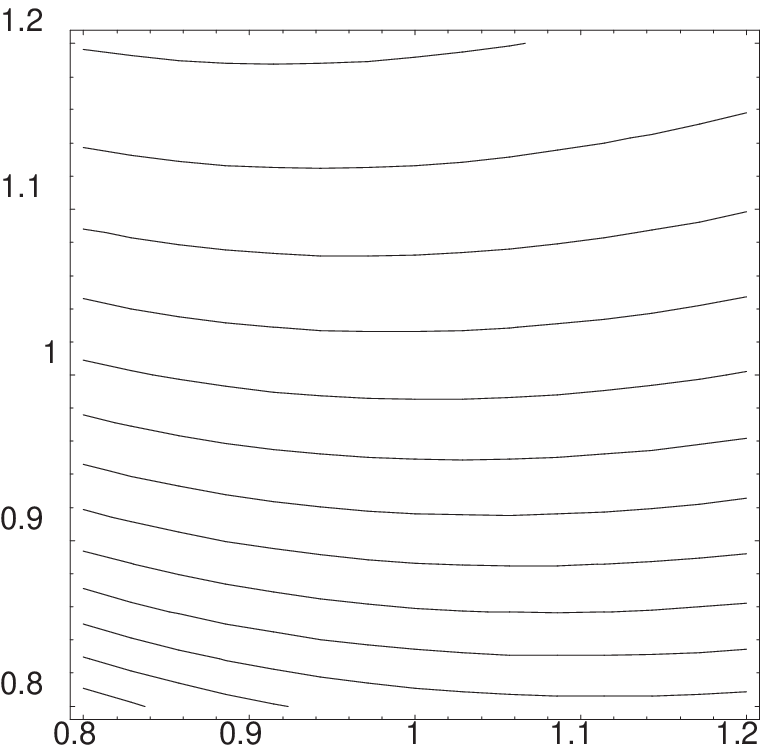} & \epsfxsize=30mm
\epsfbox[0 100 200 200]{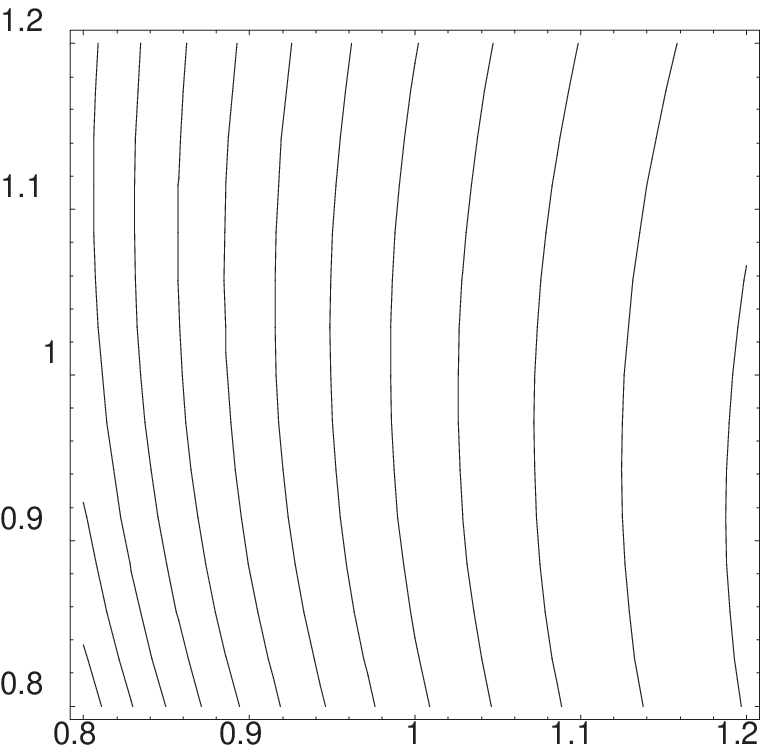} \vspace{20mm}\\
\vspace{20mm}

V & \epsfxsize=30mm\epsfbox[0 100 200 200]{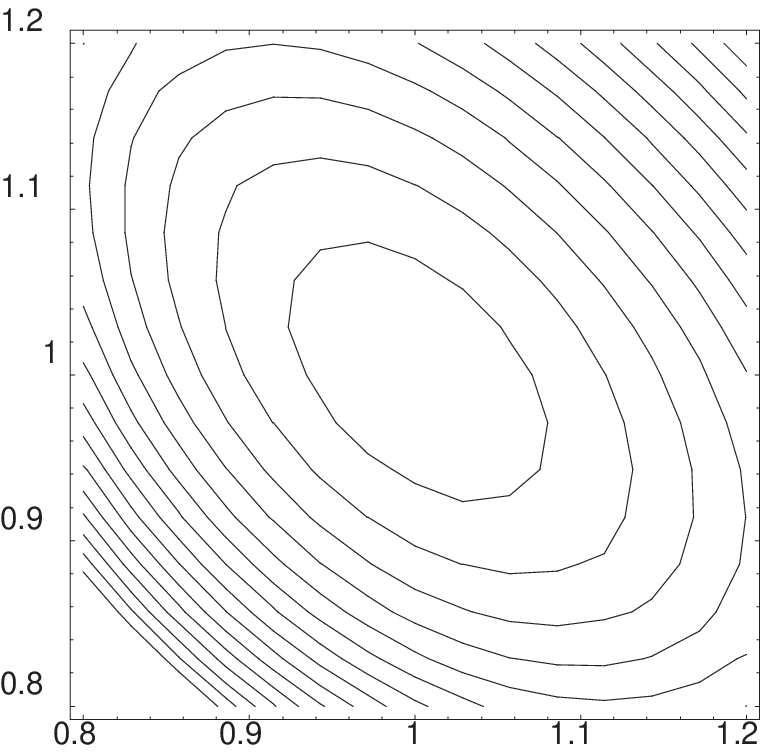} &
 \epsfxsize=30mm\epsfbox[0 100 200 200]{stuv1.eps} &
 \epsfxsize=30mm\epsfbox[0 100 200 200]{stuv1.eps} \\
\vspace{7mm}

P & \epsfxsize=30mm
\epsfbox[0 100 200 200]{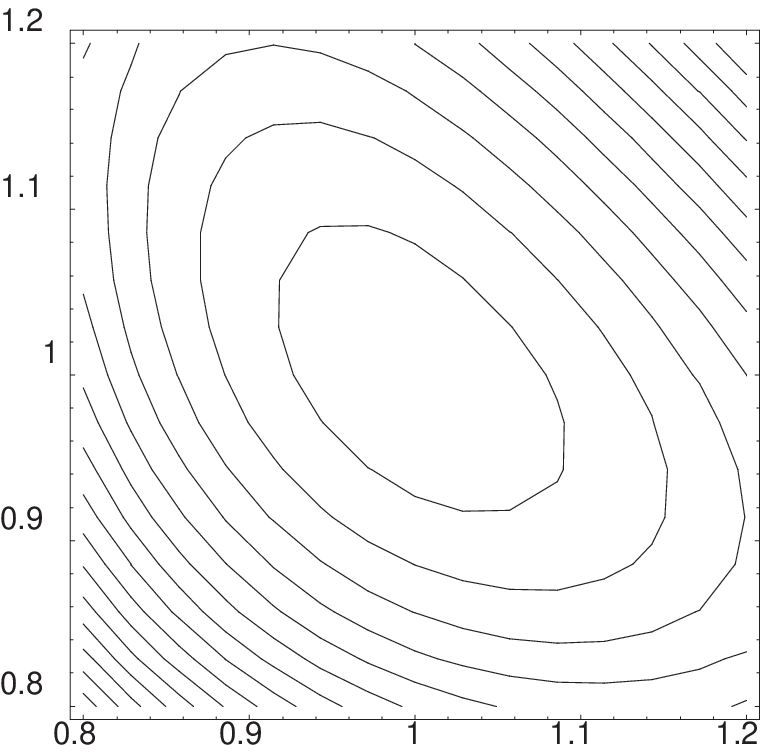} & \epsfxsize=30mm
\epsfbox[0 100 200 200]{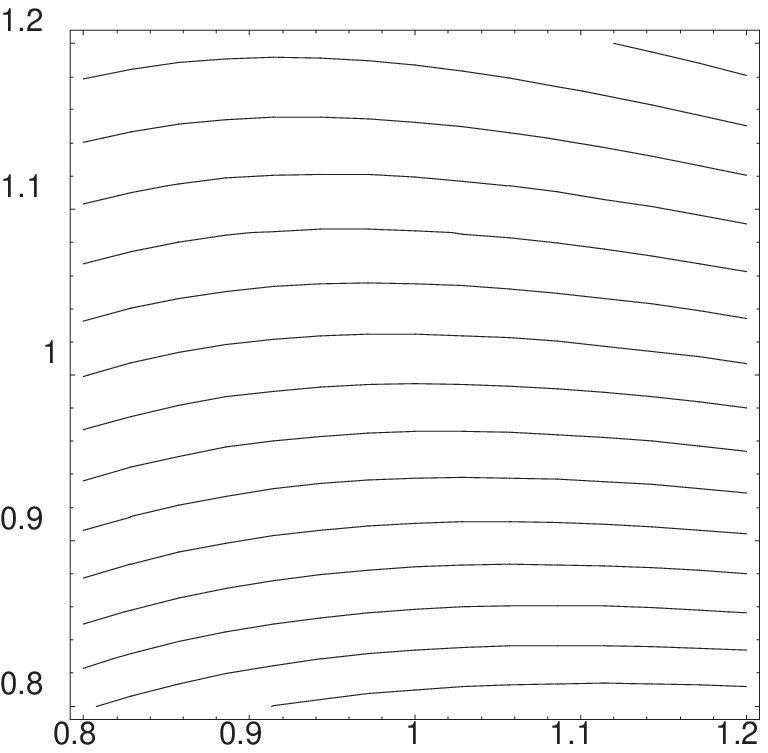} & \epsfxsize=30mm
\epsfbox[0 100 200 200]{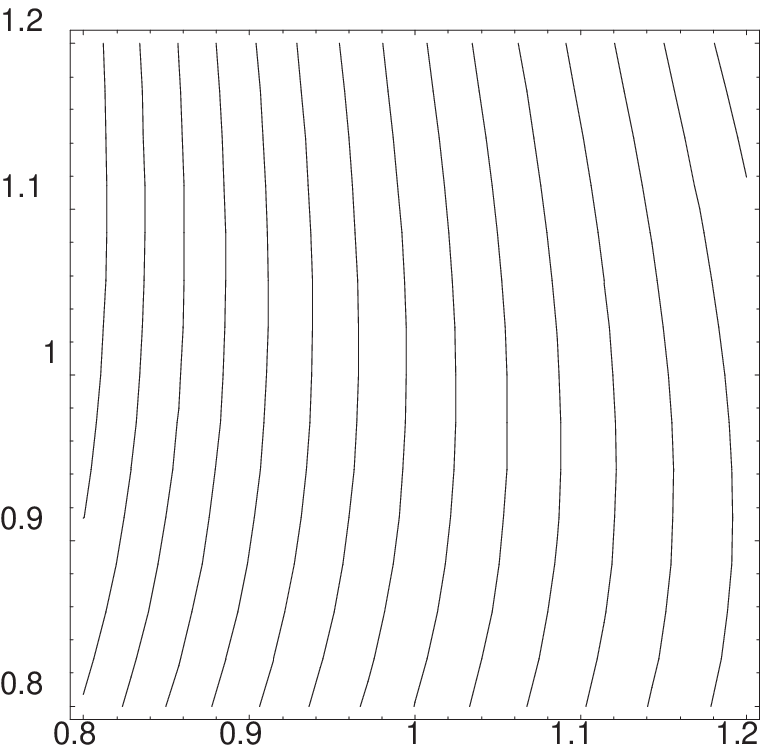} 
\end{array}
$$
\medskip
\caption{Contour plots of $Z^2$, $V$ and $P$ for charge configurations
$(q_S,q_T,q_U)=(1,1,1),(1,-1,1),(1,-1,-1)$ }
\end{figure}
Hence, we see that the
behavior of $Z$ and $V$ can be quite different from those of supersymmetric
extreme black holes.

The  gauged central charge 
 potential $P$ does admit also additional critical
points for special charge configurations.
If only one of the three charges is non-vanishing, 
$Z$ and $V$ do not have any critical points
at all. However, $P$ is {\em identically}
zero in this case. This  has been already observed in
\cite{Gunaydingeometry}.

\section{Boundaries of the K\"ahler  and  Extended K\"ahler Cones}
In the previous sections we have concentrated  on a
particular
version of $d=5, N=2$ supergravity with a given set of intersection
numbers and have discussed critical points of
the BPS mass, $V$ and $P$ respectively.
Special geometry by itself did not require the moduli
fields necessarily to be positive, as long as the
gauge couplings and the moduli space metric are regular
and positive definite. However, since these moduli 
correspond to  sizes of 2-cycles in the Calabi-Yau
 three-fold they must be positive, thus defining the K\"ahler
cone. If one of the moduli
approaches zero one of the following three scenarios  takes
place \cite{wittenM}:

(1) a complex curve $E$ (2-cycle) collapses as one approaches the
boundary. This results in a topology change via a flop
transition between two different but birationally equivalent 
 $CY_3$ manifolds.

(2) a complex divisor $D$ may collapse   (a) to a curve, or  (b) to  
a point.

We  first recapitulate some  essential features of the 
flop transition. Detailed discussions can be found in
\cite{wittenM, Antoniadisfive}.
We start with a cubic surface defining an effective $d=5$ supergravity
before the phase transition
\begin{equation}
{\cal V}  = C_{IJK} t^I t^J t^K  \ 
\label{before}\end{equation}
where all $t^I$ are positive.
Let the $t^2=0$ boundary of the K\"ahler cone constitute a flop
transition. Generically, the gauge couplings $G_{IJ}$ and the moduli space  
metric $g_{ij}$
are regular and non-degenerate at the flop transition line.

A phase transition from a positive to a negative $t^2$ value has the
following geometrical interpretation: a 2-cycle shrinks
to a zero size and  is subsequently blown up  into a topologically
distinct  manner. 
The size of the new 2-cycle is given by $t'^2=-t^2$.
Mathematically, the effect is that the topological 
intersection numbers change
 by a new term $C_{222}=-\frac{1}{6}$ \cite{Antoniadisfive}. 
In \cite{wittenM}
a physical interpretation of this topology change via  one-loop
renormalization was given.  At the $t^2 = 0$ transition point, 
 BPS states in a massive hypermultiplet
with charge $q=(0,\pm1,0..,0)$ and mass $m=t^2$ becomes massless.
The fermions in this multiplet  contribute to the parity-violating
one-loop diagram involving three $A^2$ photon vertices. 
 This one-loop diagram gives rise to an induced Chern-Simons
interaction whose coupling is governed by the intersection number 
$C_{222}$.
Across the transition the mass parameter $m$ becomes 
formally negative.  The induced Chern-Simons coupling is proportional
to the sign change of these mass parameter. Once the numerical
factors and multiplicity factor two are properly
 taken into account, it turns out that the 
parity-violating one-loop Feynman diagram induces a change of the
topological
intersection numbers of $\Delta C_{222} = - 1/6$.
 By a supersymmetry non-renormalization 
theorem, there is no further perturbative nor non-perturbative
renormalization  to this result beyond one-loop.
 The prepotential of the new phase, when expressed in variables
of the original CY space, is given by
\begin{equation}
\hat {\cal V}  = C_{IJK} t^I t^J t^K  -{1\over 6}  t^2 t^2 t^2  \ .
\label{after}\end{equation}
In this form $t^2$ is negative, hence, while
satisfactory for the original effective supergravity, this description
is not satisfactory for the flopped CY
manifold. We need to switch to different  coordinates so that
the new K\"ahler cone past the phase transition is well described.
After a  change of variables, hence, a change of basis for 
basic cycles in the new phase, of the type
\begin{equation}
t^2= - t'^2 \qquad t^1 = t'^1 + t^2 \quad  etc.
\end{equation}
one finds  that the new prepotential
\begin{equation}
{\cal V'}  = C'_{IJK} t'^I t'^J t'^K  
\label{prepp}
\end{equation}
describes the new K\"ahler cone for the flopped but 
birationally equivalent $CY_3$ space where all intersection numbers
$C'_{IJK}$ and all moduli $t'^I$
are positive.  

The extended K\"ahler cone consists of the union of  K\"ahler cones  
related by flop transitions. This enlarged region has boundaries 
 \cite{wittenM}  at which divisors shrink to
a complex curve (2a) or a point (2b). In both cases the
moduli space simply ends as opposed to four dimensional cases, where
non-geometric conformal field theory phases exist 
beyond the extended
K\"ahler cone.

In case (2a) the tension of  a BPS string is proportional to the
magnetic central charge $Z_m$ and  behaves as 
\be
Z_m\sim (t^k)^1
\ee
where $t^k$ denotes the shrinking 2-cycle.   Associated
with a shrinking 2-cycle is an $A_1$ singularity. Therefore, 
as approaching the type (2a) boundary of the extended K\"ahler cone,
we expect a $SU(2)$ 
gauge symmetry enhancement. 

The physics at the (2b) transition is more involved.
Here,  the BPS strings become tensionless with
\be
Z_m\sim (t^k)^2.
\ee
This shows that excitations
 of this magnetic string become massless at
the same rate as electric point-particle  BPS states. 
Extremal transitions are possible here if the singularity left by the
collapsing divisor can be removed by the appearance of new
complex structure deformations. As in the $d=4$ case, K\"ahler deformations
can be traded  for complex structure deformations.

We will not attempt to give a more detailed discussion of
the general framework here. We are mainly interested in
the behavior of various physical quantities
across the boundary of flop phase transitions.

\subsection{Physical Quantities at the Flop Transition}
The natural question to address is how the electric central charge $Z$, the magnetic string tension $Z_m$  and the other
potentials are affected by these phase transitions, particularly by
the flop phase transition.
We will study  the  analyticity properties of various physical
quantities at this transition line. Quantities of particular interest are

\noindent
1. the electric central charge $Z=t^I q_I$ for particle states,

\noindent
2. the gauge couplings $G_{IJ}$,

\noindent
3. the black hole potential $V= \frac{6^{\frac{1}{3}}}{4}G^{IJ} q_I q_J$,
and

\noindent
4. the magnetic central  charge $Z_m=t_I m^I$ which determines the tension
of BPS string states.

For this purpose we describe the system before and after  
the transition using the same set of special coordinates $t^I$.
At the transition, one of the independent special coordinates
(for concreteness $t^2$) changes its sign.
Let the prepotential for positive $t^2$ be given by
\be
{\cal V}( t^1, \dots ,t^{n-1}, t^n)  \la{prepold}
\ee
where $t^n$ is the function of $t^1,...,t^{n-1}$ defined by
$\Nu=1$. Then the change in
topology induces a change in the prepotential for the region of 
negative $t^2$, corresponding to the induced change in the 
Chern-Simons coupling
$C_{222}$. The prepotential in this K\"ahler cone is given by
\be
\hat\Nu={\cal V}( t^1, \dots ,t^{n-1}, \hat t^n)-\frac{1}{6}t^2 t^2 t^2.
\la{prepnew}
\ee
The notation $\hat t^n$ is to emphasize the fact that the new term in
the prepotential will change the functional form of the dependent variable
as a function of the independent ones. 

To understand the behavior of various potentials
we first need to work out how $t^n$ is affected by the transition,
and how the charges in both CY spaces have to be identified.
Let us first consider $t^n$.
Close to the transition, the change enforced by the constraints
$\Nu=1$ and $\hat \Nu =1$ in $t^n$ is easily found from
(\ref{prepold}) and (\ref{prepnew}) to be
\begin{equation}
\Delta t^n = \hat t^n - t^n \sim   
-{1\over 6} (t^2)^3\left(  
{\partial
{\cal V} \over \partial t^n}\right)^{-1} \ \qquad {\rm as} \
t^2\rightarrow 0.
\ee
The prepotential ${\cal V}$ is given by a function  
which may have a
 cubic, quadratic, linear and constant dependence on $t^n$. 
Therefore in  generic cases
the function $\left( {\partial {\cal V} \over \partial  
t^n}\right)^{-1}
$ is regular near the boundary.  The case  that the  
dependence on $t^n$
enters only via  $t^n t^2 t^I$ or  $t^n t^2 t^2$ is excluded since this  
would lead
to $G_{nI}\rightarrow 0$ and
 divergent gauge couplings at
the phase transition. This would not constitute a flop transition.
Thus, we have established the fact that the dependent variable is affected
only very mildly, as it behaves as 
\begin{equation}
\Delta t^n = \hat t^n - t^n \sim  (t^2)^3  \rightarrow 0  \ , 
\qquad {\rm as}
\ t^2 \la{tnbeh}
\rightarrow 0.
\end{equation}

The question of the identification of the 
charges in both Calabi-Yau
manifolds also finds an easy answer. As the charges $q_1,q_3,..,q_n$
correspond to membrane wrappings around two-cycles which do
not shrink to zero at $t^2=0$, they clearly cannot change under
the transition. Slightly more subtle is the issue of $q_2$.
We will adopt the supergravity point of view \cite{wittenM}
and note that the topology change
gives rise to a change to
 the one-loop quantum effect of massive states.
Besides the discrete shift of the Chern-Simons coupling the transition
is perfectly smooth. We thus conclude
that $q_2$ is unaffected as well by the flop
if $t^2$ is taken to be negative after the transition. If one  
 adopts the natural Calabi-Yau variables in this flopped side also, 
implying
\be
t^2\rightarrow \hat t^2=-t^2
\ee
then $q_2$ also changes its sign correspondingly\footnote{Generically there
are more redefinitions necessary which also affect the charges.}.
However, we stress that in the variables of the
original Calabi-Yau space the charges should be unchanged.

With these preliminaries it is straightforward to determine the analyticity
properties of $Z$, as the central charge
$ \sum_1^{n}(t^I q_I)$ is affected by the topology change only via the
induced change in  $t^n$.
 Before and after the transition we have
\begin{equation}
Z= \sum_1^{n-1}( t^I q_I) + t^n (t^1, \dots t^{n-1})q_n \qquad \hat Z=
\sum_1^{n-1}( t^I q_I) + \hat t^n (t^1, \dots t^{n-1})q_n
\end{equation}
and therefore
\begin{equation}
\Delta Z = \hat Z -Z = ( \hat t^n -t^n) q_n \sim (t^2)^3 q_n  
\rightarrow 0
\ , \qquad {\rm as} \ t^2 \rightarrow 0.
\end{equation}
Thus the electric central charge for a generic set of charges is continuous  
across the phase
transition.  This is also the case for
the first and second  derivatives over $t^2$.  
The third
derivative of $Z$  is discontinuous in general.
This point is of immense importance, as it implies that there
is  {\it at most} one supersymmetric stabilization point throughout the
entire extended K\"ahler cone! As $Z$ and its first derivative
are continuous at flop transitions, additional extrema would
require at least one maximum or saddle point. However, special 
geometry is valid in any section of the extended cone, hence
any extrema must be a minimum. 
This proves  {\it uniqueness} of the supersymmetric
stabilization throughout the entire extended K\"ahler cone.

 This important observation bears the following implications. 
If a charge
configuration is realized by a black hole in space-time the moduli
fields at the horizon will take the values of the unique critical point. 
However, their asymptotic values can be taken arbitrarily, in particular,
they can be chosen to the values of a different K\"ahler cone from the
K\"ahler cone in which the near-horizon values are taken. It is clear that
then the Calabi-Yau space at infinity is topologically different from
the space close to the black hole. In this sense, the $d=5$ BPS black hole
induces an interpolation of CY spaces of different
topology.
Surrounding the black hole is a surface of codimension one at which the
the topology of the internal CY 
space changes, and which has additional massless states localized
on it.
This surface around the BPS black hole constitutes a
$d=4$ domain wall world with distinct massless physical spectrum.

We now turn to the gauge couplings $G_{IJ}$.
They are defined by the second derivative of the prepotential
$G_{IJ}=-{1\over {2 \cdot 6^{\frac{2}{3}}}}
\partial_I \partial_ J \ln {\cal V}|_{{\cal V}=1}$. To lowest order in
$t^2$ they change at the transition line as
\begin{equation}
\Delta G_{IJ} =  {1\over {2\cdot 6^{\frac{2}{3}}}}  
\delta _I^2 \delta_J^2 t^2 \rightarrow 0
\ , \qquad {\rm as} \ \ t^2 \rightarrow 0. \la{Gbeh}
\end{equation}
The first derivative of $G_{22}$ is discontinuous
since
\begin{equation}
\partial_{t^2}  \hat  G_{IJ}  - \partial_{t^2}   G_{IJ} = 
{1\over {2\cdot 6^{\frac{2}{3}}}}  \delta_I^2 \delta_J^2
\ , \qquad {\rm as} \ t^2 \rightarrow 0.
\end{equation}
The continuity and analyticity  of the black hole 
potential $V=\frac{6^{\frac{1}{3}}}{4} 
q_I G^{IJ}  q_J$  
is determined by the properties of the gauge couplings.
With (\ref{Gbeh}) we find
\begin{equation}
\hat V - V =  \frac{6^{\frac{1}{3}}}{4}q_I ( \hat  G^{IJ}  -   G^{IJ})q_J  =
 -  \frac{1}{8 \cdot 6^{\frac{1}{3}}}  t^2 \:
(q_I    G^{I2} )^2   \rightarrow 0
\ , \qquad t^2 \rightarrow 0.
\end{equation}
Note that $V$  always has an upward concave kink around
the transition, since
\begin{equation}
\partial_{t^2}  \hat V -  \partial_{t^2}  V =   -\frac{1}{8 
\cdot 6^{\frac{1}{3}}}   (q_I   
  G^{I2}
)^2\leq0
\ , \qquad  {\rm as} \ t^2  \rightarrow 0.
\end{equation}

This observation might turn out to be crucial for
proving  uniqueness of non-supersymmetric stabilization as well.
Unfortunately, the nature of non-supersymmetric extrema is
not quite  understood yet.
 On the other hand,
 if the uniqueness of the critical point of $V$ can be 
shown, then the immediate conclusion is that the black hole
entropy is not affected by the topology change to a 
birationally
equivalent $CY_3$ at infinity. This is also what one would expect.
Clearly,  the gauged central charge potential
$P$ has the same analyticity as  the black-hole potential
$V$, as it is a linear combination of $Z$ and $V$.

Finally, it is very easy to study the analyticity properties
of the BPS string tension 
\be
Z_m=t_I m^I.
\ee
As we also do not expect the string charges to be affected
by the flop transition, any discontinuity in $Z_m$ can only
enter via the $t_I$ which are given by
\be
t_I= 4 \times 6^{1/3} G_{IJ} t^J.
\ee
Hence we find with (\ref{Gbeh}) and (\ref{tnbeh}) 
\be
\Delta Z_m \sim m^2 (t^2)^2.
\ee 
In conclusion, the BPS tension and its first derivative
are continuous across the flop transition, the
second derivative generically is not. Nevertheless, the
smoothness is sufficient to show the uniqueness also of
the stabilization point of $Z_m$ in the extended
K\"ahler cone.

\section{Examples of Calabi-Yau Compactifications}

We now give an explicit example for the discussions
above. We study an extended K\"ahler cone (displayed in
Figure 2) consisting of two elliptically
fibered $CY_3$ spaces with base $F_1$ but with different
triangulations of the toric diagram. 
The first $F_1$ vacuum
is denoted as model III and contains three K\"ahler moduli.
Its second Chern class is characterized by
\be
c_2(J_I)=\int c_2\wedge J_I=(92,36,24).
\ee
In this space a flop can be performed to the other
triangulation of the base $F_1$ (model II), leading to
different intersection numbers and
\be
c_2(J_I)=(92,102,36).
\ee
Here, a complex divisor can be shrunk to zero size
and the resulting singularity can be resolved by blowing up
29 three-cycles leading to an elliptic fibration with base $P_2$,
model I.
This space contains only two K\"ahler moduli and has
\be
c_2(J_I)=(102,36).
\ee
These $CY_3$ manifolds and their interrelation  have been discussed
extensively  in \cite{Louis,Morrison,Candelas}.

We will explicitly study the stabilization of the  central charge,
$Z$,
the black hole potential $V$ 
and the gauged  central charge potential
$P$ in all regions
of the extended K\"ahler cone and analyze how various
potentials behave under the flop  transition.
 The main results of our study are as follows. We find
 new critical points for the
gauged central charge $P$ which were out of reach of the
discussion in \cite{Gunaydingeometry}. These extrema can
be saddle points or maxima, as opposed to the critical points
which coincide with the extrema of $Z$, as those are all minima.
The supersymmetric stabilization equations are explicitly 
solved in all regions of the moduli space and we will graphically
demonstrate the smoothness of $Z$ and the kink in $V$ along the
flop transition.
We also study the tension of  the BPS strings at
the boundaries of each K\"ahler cone
and find that our results
are in agreement with the general considerations of \cite{wittenM}.

For convenience we start the discussion with the $P_2$ vacuum, then blow 
up a four-cycle and finally consider the flopped vacuum.

\begin{figure}
\centerline{\epsfysize=17cm \epsfbox{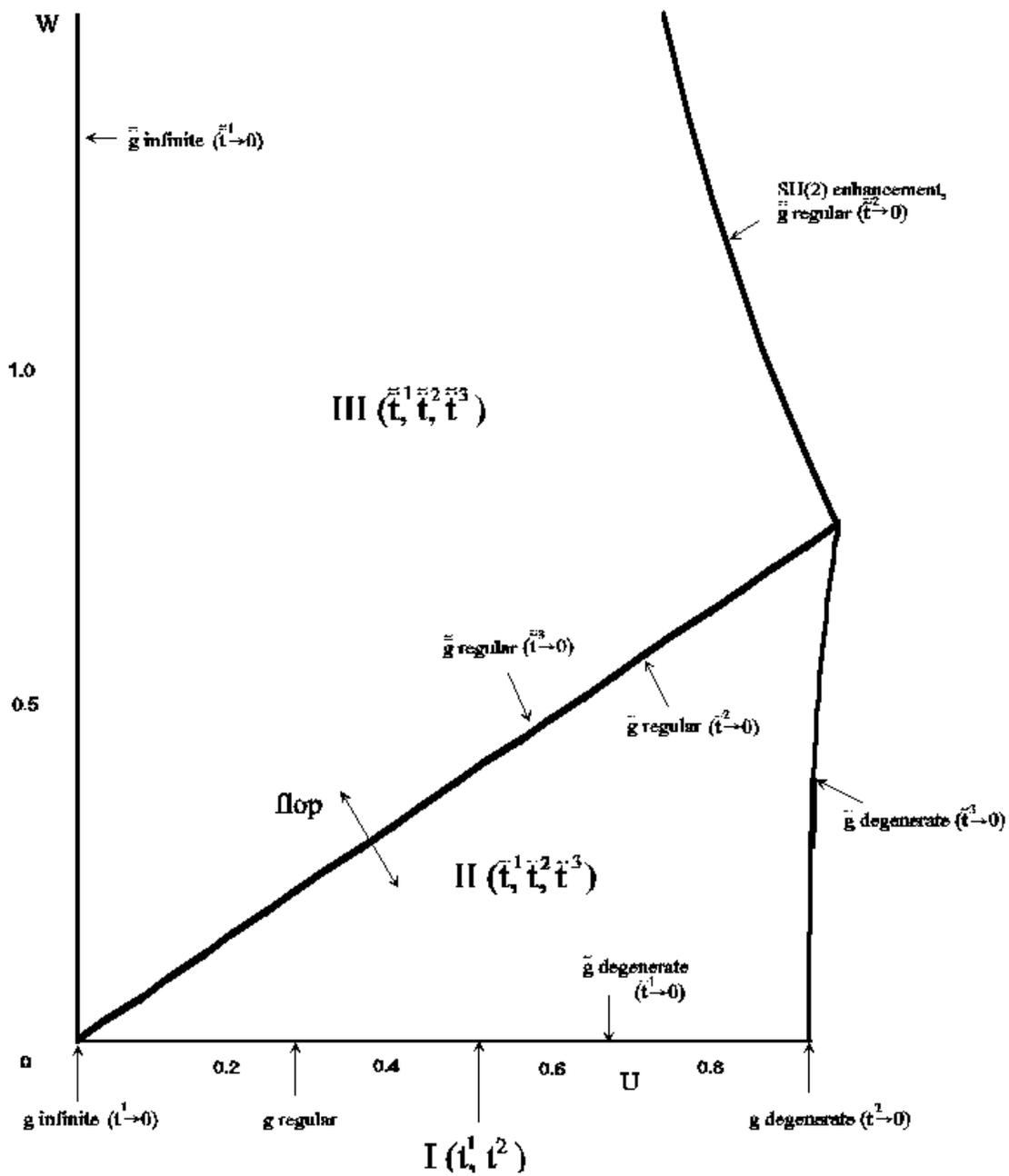}}
\bigskip
\caption{The extended moduli space}
\end{figure}

\subsection{Model I: the Base $P_2$ Vacuum}

This Calabi-Yau compactification corresponds to the lower
boundary of Figure 2. It is a rather simple example
as it admits only one independent 
moduli field. Yet, it incorporates rich and surprising
structures.

Its prepotential is given by
\be
{\cal V}_I=9(t^1)^3+9 (t^1)^2 t^2+3 t^1 (t^2)^2=1.
\ee
To achieve a consistent notation for all three 
connected models  
we perform a coordinate redefinition to variables
$U$ and $T$, related to the basic cycles $t^1$ and $t^2$ by
\bea
U&=&6^{\frac{1}{3}}t^1 \nn \\
T&=&6^{\frac{1}{3}}(t^2+\frac{3}{2}t^1). 
\eea
The constraint on the moduli fields translates  to
\be
{\cal V}_I=\frac{3}{8}U^3+\frac{1}{2}U T^2=1. \la{precandTU}
\ee
For convenience we choose  $U$ as an independent variable. In this
case, $T$ is given by
\be
T=\sqrt{\frac{2}{U}(1-\frac{3}{8}U^3)}.
\la{TUcand}
\ee
Note that  it is  the basic cycles $t^1,t^2$ 
of the Calabi-Yau manifold  and not the $T, \,\, U$ variables
which should be restricted  to be positive.

To understand the structure of the moduli space one
has to compute the metric which in this case is a
scalar.
It is  determined to be
\be 
g=\frac{6-9U^3}{8U^2-3U^5}.
\ee
Important are also the gauge couplings. 
We do not give the explicit form of $G_{IJ}$ here 
but note that its determinant is given
by
\be
\det G_{IJ}=\frac{1}{24 \times 6^{1/3} U}(1-\frac{3}{2} U^3).
\ee
In the region $0\leq t^1, t^2$, 
$g$ and $\det G$ are regular as expected. 
They diverge  at $t^1\sim U=0$ and degenerate at  
\be
U_0=(\frac{2}{3}) ^{1/3}.
\ee
This is also  the  point where $t^2$ turns negative.
Hence,
$0\leq U \leq U_0$ defines the  domain of validity. 
The regularity of $T$ in this region and on
 the boundary $t^2=0$  is quite crucial as
it implies that $Z$ and $\partial Z$ are regular here also.

These considerations allow us to predict some properties
of the potentials. First of all, for a range of charge configurations
$(q_U,q_T)$ 
we do expect 
a critical point
of the central charge in the regular domain which in turn
is  a minimum of the potentials $V$ and $P$. 

Since $T$ diverges
at $U\rightarrow 0$ all three potentials diverge to positive infinity
at this boundary of the moduli space. On the other side of the
domain of validity, $Z$ and
$\partial Z$ are regular but $g^{-1}\rightarrow \infty$, hence
$V$ generically diverges to positive infinity whereas $P$ turns negative
towards  infinity. This implies a second new critical
point for potentials of gauged supergravities\footnote{It is
well known that the stabilization issue of critical points in
gauged supergravities is rather subtle. The stabilization
of this new critical point remains to be analyzed.}! In fact, this
extrema must be a maximum of $P$ (implying a minimum of the
gauge  potential, as we flipped the sign for convenience).

The second prediction we can make is that $V$  {\it always}
admits a minimum, as on both boundaries it diverges to positive
infinity.

Let us verify our general considerations. First we solve the stabilization
equations, leading to the critical points of the
central charge. 
The  charges are denoted by $(q_U,q_T)$ where we take $q_U\geq 0$
without loss of generality, since $(q_U,q_W)\rightarrow -(q_U,q_W)$
leaves the potentials invariant. In the domain of validity
we find the solution
\bea
U_{\rm crit}&=&\frac{2}{\sqrt{3 Z^{\rm crit}}}
\sqrt{q_U-\sqrt{q_U^2-\frac{9}{4}q_T^2}} \nn \\
T_{\rm crit}&=&\sqrt{\frac{3}{Z^{\rm crit}}}
\sqrt{q_U+\sqrt{q_U^2-\frac{9}{4}q_T^2}} 
\la{stabcand}
\eea
with 
the central charge 
\be
Z^{\rm crit}=
\left[q_U\frac{2}{\sqrt{3}}
\sqrt{q_U-\sqrt{q_U^2-\frac{9}{4}q_T^2}}
+q_T
\sqrt{3 q_U+3 \sqrt{q_U^2-\frac{9}{4}q_T^2}}\right]^{2/3}.
\ee
The charge configurations allowing for a stabilization
of the central charge are constrained
by
\be
q_U\geq \frac{3}{2} q_T\geq0.
\la{candrest}
\ee
As anticipated, in the domain $g>0$ ($t^1,t^2>0$) we 
find precisely one extremum of the central charge (\ref{candrest}).
This should coincide with the extremum of the potentials according to
our considerations of the last section. 
We will not give the explicit form of $V$ and $P$ here as functions
of $U$ but rather 
consider the specific charge configurations $q_U=4, q_T=2$
and $q_U=2, q_T=2$. The results are plotted in Figure 3.
This can be done without much  loss of generality as the
general structure of the potentials is rather insensitive to a
 specific choice of charges, as long as we avoid the boundary
values $q_T=\frac{2}{3}q_U$ and $q_T=0$.

\begin{figure}
\centerline{\epsfxsize=80mm\epsfbox{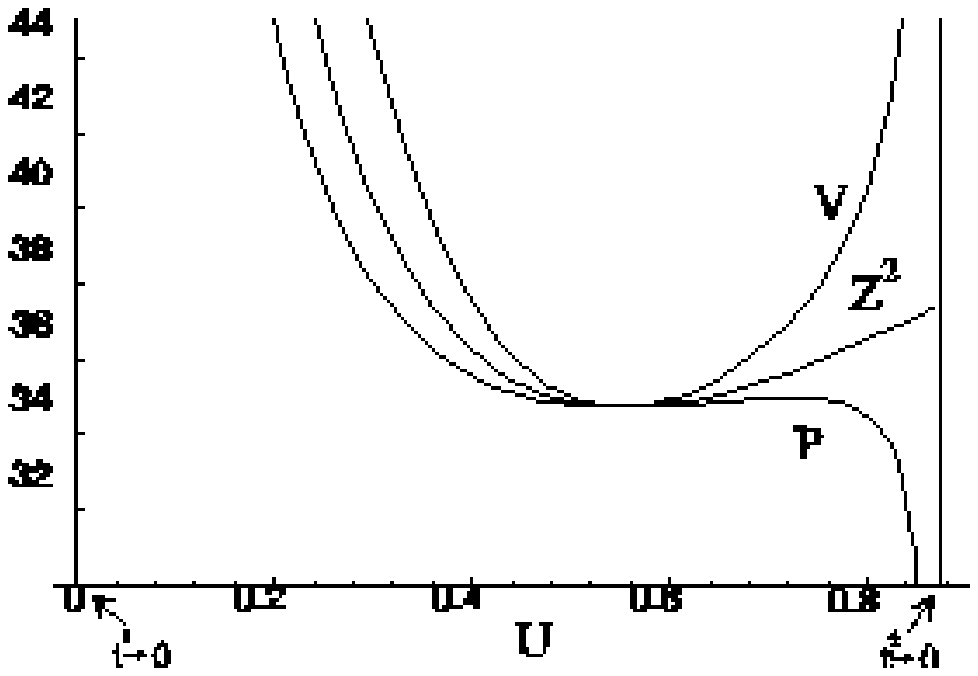}\epsfxsize=80mm
\epsfbox{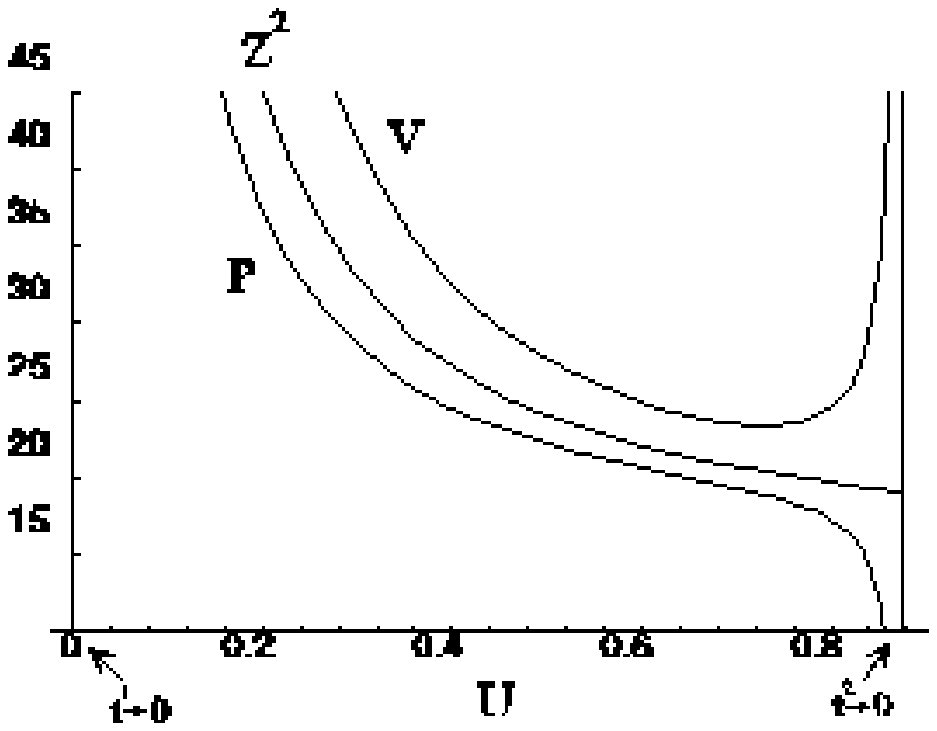}}
\medskip
\caption{$Z^2$, $V$ and $P$ for $(q_U,q_W)=(4,2)$ (left) and
$(q_U,q_T)= (2,2)$ (right)}
\end{figure}

For the first set of charges the picture is
as expected.
The minima of  $Z^2,P$ and $V$
coincide exactly at the position predicted by
(\ref{stabcand}). However,  $P$ indeed admits a
new maximum which does not correspond
to critical points of the central charge. 

The situation is very different for the configuration
$(q_U,q_T)=(2,2)$. The central
charge does not get stabilized anymore, also $P$ does
not admit  critical points. However, $V$ still has
a minimum in agreement with our discussion above.
This minimum corresponds to the stabilized values of the
moduli in the non-supersymmetric extreme
black hole solution with the same charges. 
In fact one finds that for
{\it all} charge configurations stabilization of 
the scalar potential takes place!

There
are also the two boundary points: 
\be
q_U=\frac{3}{2} q_T \ \ \ \ {\rm and}\ \ \ \  q_T=0 \la{candspecial}.
\ee
Here, we find stabilization at the boundaries 
(i.e. on the singularities) and the potentials become
finite there. An example is given in Figure 4, where
the charge configuration $(q_U,q_T)=(3,2)$ is shown
to admit stabilization at the point $U=U_0$, i.e. at $t^2=0$.
\begin{figure}
\centerline{{\epsfbox{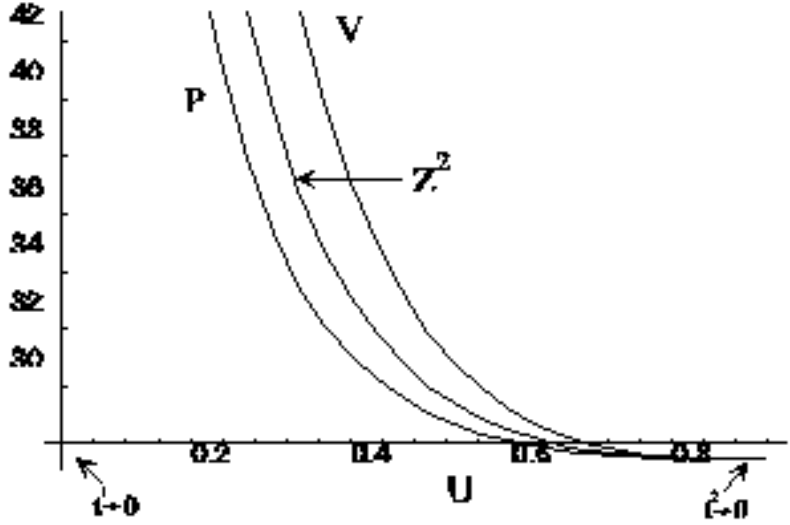}}}
\medskip
\caption{$Z^2,V,P$ for $q_U=3, \ q_T=2$}
\end{figure}
This is rather peculiar as it implies that
stabilization takes place right on the edge of a Calabi-Yau
space. Stabilization at $t^1=0$ occurs for $q_T=0$.

To summarize the discussion of
$V$ we  emphasize again that in this example for {\it all} 
charge configurations  $V$ has a minimum 
in the allowed region of moduli fields (or on its
boundaries), as opposed to $Z$ and $P$, where extrema are
 found only when (\ref{candrest}) is satisfied. 

It is known (see for example \cite{Louis}) that in this
Calabi-Yau manifold 29  complex structure moduli 
can be shrunk to 
zero size, and the resulting
singularity can be resolved by blowing up a four-cycle. In this
manner we obtain a new Calabi-Yau manifold with one more 
K\"ahler moduli
field, which will be discussed now.

\subsection{Model II: the Base $F_1$ Vacuum }

This space (model II of Figure 2) is  more 
complex than the previous one. 
Nevertheless, the general structure of critical points is 
similar to what we  have learned above.  

If a complex divisor is blown up in the $P_2$ space, 
resulting in a new
two-cycle $W$, it turns out that the new
prepotential is given by \cite{Louis}
\be
\Nu_{II} = \frac{3}{8}U^3+\frac{1}{2} U T^2 -\frac{1}{6}W^3.
\la{premar}
\ee
Clearly, in the limit $W\rightarrow 0$ one recovers
(\ref{precandTU}).

To find the boundary of this K\"ahler cone a transformation
to the basic cycles 
\bea
\tilde t^1&=&W  \nn \\
\tilde t^2&=&U-W \\
\tilde t^3&=&T-\frac{3}{2} U \nn
\eea
has to be performed in terms of  which the prepotential
becomes
\be
\Nu_{II}=\frac{4}{3}(\tilde t^1)^3+\frac{3}{2}(\tilde t^2)^3
+\frac{9}{2} (\tilde t^1)^2 \tilde t^2 +\frac{9}{2}
\tilde t^1 (\tilde t^2)^2 +\frac{3}{2}
(\tilde t^1)^2 \tilde  t^3+\frac{3}{2}
(\tilde t^2)^2 \tilde t^3+\frac{1}{2}
\tilde t^1 (\tilde t^3)^2+\frac{1}{2}
\tilde t^2 (\tilde t^3)^2+3 \tilde t^1 
\tilde t^2 \tilde t^3,
\la{premart}
\ee
and the domain  of this K\"ahler cone is then defined
by $\tilde t^I\geq 0$.

%
%
%

For the discussion of the critical points the variables
$(U,T,W)$ are more convenient.
 If we choose $U$ and $W$ as independent variables, the constraint
$\Nu_{II}=1$  leads to
\be
T=\sqrt{\frac{24 -9 U^3 +4 W^3}{12 U}}
\ee
and the determinant of the moduli space metric is given by
\be
\det \tilde g_{ij}=\frac{3 W(6 -9 U^3 + W^3)}
{2 U^2 (24 -9 U^3 +4 W^3)}.
\ee
The determinant of the gauge couplings is found to be
\be
\det \tilde G_{IJ}=\frac{W}{288 U}-\frac{U^2 W}{192}+\frac{W^4}{1728 U}.
\ee
The boundary of the K\"ahler
cone is reached by setting one (or more) of the fundamental
cycles to zero. At these points one can have a flop 
transition or reaches the end of the extended
K\"ahler cone. At $\tilde t^1=W=0$ 
(lower boundary in Fig. 2) we find $\det \tilde g_{ij}=0$ 
and we recover
model I as mentioned before. The 
moduli space metric also degenerates at the boundary
$\tilde t^3=T-\frac{3}{2}U=0$ (right boundary in Fig. 2). 
We note that some  gauge fields become
strongly coupled at these boundaries. 

$\tilde t^2=U-W\rightarrow 0$ (upper left boundary) turns out to be  a 
 flop transition.
Here, the metrics are completely regular and  very 
special geometry is still valid if we 
pass this boundary. However, physically we enter the new Calabi-Yau space
of model III with different intersection numbers. 

We begin the discussion of critical points with a few
general observations.
If we choose a charge configuration which allows for
a stabilization of the central charge 
then the arguments given for model I  do
apply again and we can deduce some of the properties of the
potentials. Consider first $P$. At each of the two intersecting 
$\det \tilde g=0$ lines (i.e. $\tilde t^1=0$ and $\tilde t^3 =0$) 
$P$ is expected to diverge to negative infinity. 
As the supersymmetric stabilization
point must be  a minimum, there ought to be three additional  critical
points: two saddle points and one maximum. Hence, we learn once
more that the structure of critical points of gauged supergravities
is rather rich. 
We also expect that 
$V$ has new critical points for charge
configurations which do not stabilize $Z$ in the valid domain.

As for model I we verify our intuition by 
solving the stabilization equations and study the
potentials with  specific charge configurations.
We find that the extremum of the central charge in the
valid region is located
at
\bea
U_{\rm crit}&=&\frac{2}{\sqrt{3 Z^{\rm crit}}}
\sqrt{q_U-\sqrt{q_U^2-\frac{9}{4}q_T^2}} \nn \\
T_{\rm crit}&=&\sqrt{\frac{3}{Z^{\rm crit}}
\sqrt{q_U+\sqrt{q_U^2-\frac{9}{4}q_T^2}}} \\
W_{\rm crit}&=& \sqrt{-\frac{6}{Z^{\rm crit}} q_W} \nn 
\eea
with the critical value of the 
central charge 
\be
Z^{\rm crit}=\left(q_U\sqrt{\frac{4}{3}(q_U-
\sqrt{q_U^2-\frac{9}{4} q_T^2})}
+q_T\sqrt{3(q_U+ 
\sqrt{q_U^2-\frac{9}{4} q_T^2})}+q_W\sqrt{-6 q_W}
\right)^{\frac{2}{3}}.
\ee
Here, $q_U>0$ was chosen without loss of generality due to
the ${\bf Z}_2$ ($q^I\rightarrow -q^I$) symmetry of the potentials.
Supersymmetric stabilization occurs only for the region
\bea
q_U&\geq& \frac{3}{2}q_T\geq 0 \nn \\
0&\geq&
q_W\geq -\frac{2}{9}\left(q_U-\sqrt{q_U^2-\frac{9}{4}q_T^2}\right).
\eea
At the  critical values (saturation of above inequalities)
$Z$ gets stabilized on the boundaries of the moduli space.
To be explicit, we find for $q_U=\frac{3}{2} q_T$ stabilization at
$\tilde t^3=0$ and for $q_W=0$ at $\tilde t^1=0$. The  latter
observation
is very much consistent with the expectation that absence of
$q_W$ drives the internal space in the vicinity of the black hole
to the $CY$ space with $P_2$ as a base.

If 
\be
q_W< -\frac{2}{9}\left(q_U-\sqrt{q_U^2-\frac{9}{4}q_T^2}\right)
\ee
no stabilization
occurs in the present K\"ahler cone. In fact, what we observe is that
the critical point moves over the boundary  $\tilde t^2=U-W=0$ into the
adjacent Calabi-Yau space.

Let us now turn to the study of $P$ via a specific example.
As a typical case with supersymmetric stabilization we consider
the charge configuration
\be
q_U=5, \quad q_T=2, \quad q_W=-\frac{1}{6},
\ee
as general properties of the potential are independent
of a specific choice. Figure 5 shows beautifully the
 structure of $P$. This potential  admits 
indeed four critical points 
as anticipated. 
The minimum agrees with the ones of $V$ and $Z^2$, the
two saddle points and the maximum are new.
\begin{figure}
\centerline{\epsfysize=9.5cm \epsfbox{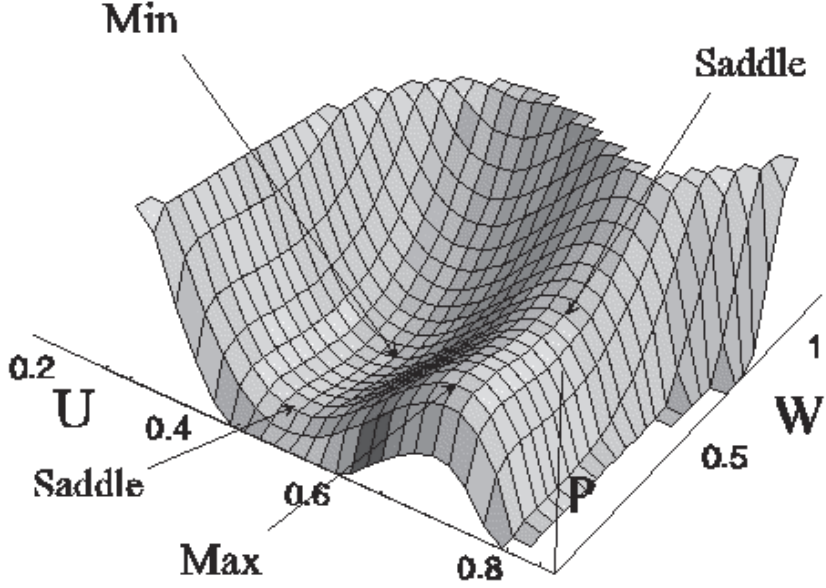}}
\medskip
\caption{Potential $P$ with 4 critical points}
\end{figure}

If $q_W$ approaches $0$, the pairs of critical points merge on the
$W=0$ line rendering the results of model I. If 
$q_W$ becomes positive, the critical points of $P$ and $Z$ 
disappear, however, $V$ still stabilizes,
showing the same characteristics as in the last example. As long
as $q_W$ is below a certain value, the black hole potential stabilizes
in the allowed domain of this Calabi-Yau. There is some
critical value for which also the non-supersymmetric stabilization of 
$V$ takes place in the adjoining K\"ahler cone in very much the same way 
as we already observed for $Z$.
Note that 
$\tilde t^2=U-W$ is the only boundary where the moduli space metric is 
non-degenerate, hence stabilization points can smoothly ``move''
across this transition line. 

Let us also comment  on the other two boundaries. In particular, we will
consider the tension of solitonic  BPS
strings.
The tension of these solitonic objects is given by
\be
Z_m=\tilde t_I \tilde m^I
\ee
where the $\tilde m^I$ are the winding numbers of  eleven-dimensional
five-branes wrapping around the four-cycles $\tilde t_I$ defined
in (\ref{tdual}).
At the boundary $\tilde t^1\rightarrow 0$ (transition to model I)
we find indeed tensionless strings
consistent with
the  arguments given in \cite{wittenM}.
This can be easily seen as the four-cycle moduli  are given by
\bea
\tilde t_1&=&\frac{4}{3}(\tilde t^1)^2 + 3 \tilde t^1 \tilde t^2 
+\frac{3}{2}(\tilde t^2)^2 + \tilde t^1 \tilde t^3
+\frac{1}{6}(\tilde t^3)^2 + \tilde t^2 \tilde t^3 \nn \\
\tilde t_2&=&\frac{3}{2}(\tilde t^1)^2 + 3 \tilde t^1 \tilde t^2 
+\frac{3}{2}(\tilde t^2)^2 + \tilde t^1 \tilde t^3
+\frac{1}{6}(\tilde t^3)^2 + \tilde t^2 \tilde t^3  \\
\tilde t_3&=&
\frac{1}{2}(\tilde t^1)^2 +  \tilde t^1 \tilde t^2 
+\frac{1}{2}(\tilde t^2)^2 + \frac{1}{3}\tilde t^1 \tilde t^3
+
\frac{1}{3} \tilde t^2 \tilde t^3. \nn
\eea
If we consider the cycle $\tilde t_1-\tilde t_2$ we find
\be
\tilde t_1-\tilde t_2=-\frac{1}{6}(\tilde t^1)^2
\ee
implying that  
the  BPS string states with charges 
$\tilde m=\pm (1,-1,0)$ indeed become tensionless with 
$Z\sim (\tilde t^1)^2$!
We argue that model I is reached
via this mechanism.

The boundary $\tilde t^3\rightarrow 0$ shows the same characteristics,
the moduli space metric degenerates and the string states with
charges\footnote{As in five dimensions all moduli are real all
states are bound states at threshold. Proving their existence
is typically quite involved and we do not attempt to do so.} 
\be
\tilde m =\pm(0,1,-3)
\ee
become tensionless also with tension $\sim (\tilde t^3)^2$.

As expected, at the flop transition $\tilde t^2=0$ it is straightforward to
verify that  none of the string states
becomes tensionless along the entire transition line, i.e. there
is no linear combination of $\tilde t_I$ which vanishes
here.

We will now  study the flopped Calabi-Yau space in detail.

\subsection{Model III: The Flopped $F_1$ Base}

In model II $\tilde t^2=U-W\rightarrow 0$ marks a topology change
of the mild form. The Hodge numbers are  
unaffected, but the intersection numbers $C_{IJK}$
change \cite{Antoniadisfive,wittenM}.
These topological data changes  lead to the new prepotential 
augmented by the term $-\frac{1}{6}(U-W)^3$ 
\be
{\cal V}_{III}=\frac{5}{24}U^3+\frac{1}{2}U^2 W -\frac{1}{2}
U W^2 +\frac{1}{2}T^2 U=1 .
\ee
As discussed in section  5.1  the charges
$q_U, q_W, q_T$ are in fact unchanged by this flop
transition. 

$U,W$ and $T$  
are again not the proper Calabi-Yau variables, the correct
parametrization to determine the boundary of this
space is 
\bea
 \ttt ^1 &=&U \nn \\
 \ttt ^2 &=&T-\frac{1}{2}U -W \\
 \ttt ^3 & = & W-U \nn
\eea
and $\ttt^I>0$ defines this K\"ahler cone. 
In these variables the  prepotential becomes
\be
{\cal V}_{III}=\frac{4}{3}(\ttt^1)^3+\frac{3}{2}(\ttt^1)^2 \ttt^2  +
\frac{1}{2}
\ttt^1 (\ttt^2)^2+ (\ttt^1)^2 \ttt^3 +\ttt^1 \ttt^2 \ttt^3=1.
\ee
The potentials are again analyzed in $(U,T,W)$ variables.  
Due to the constraint, $T$ is now given by
\be
T=\frac{6-8 U^3-9 U^2 W-3 U W^2}{6U(U+W)} .
\ee
We find for the moduli space metric and gauge couplings
\be
\det \gtt_{ij}=\frac{3-4 U^3}{4U^2(U+W)^2}
\ee
and
\be
\det \Gtt_{IJ}=\frac{3-4 U^3}{864}.
\ee
Hence, $\gtt $ diverges at $\ttt^1\rightarrow 0$, (left boundary 
of Fig. 2), but is
 regular at $\ttt^2\rightarrow 0$ (right boundary) and 
$\ttt^3\rightarrow 0$ (lower boundary). 
Regularity at the latter transition is no surprise, as we 
have just 
pass back into the Calabi-Yau space of above via the flop. However,
$\ttt^2\rightarrow 0$ marks the end of the moduli space in
{\it finite} distance. We will later interpret this line
as the transition where a divisor shrinks to a complex curve.
Noteworthy is that, as all metrics are regular here,
there is no protection that ensures stabilization
{\it before } the boundary, not even for $V$. 

As the behavior of the potentials is generically not very different from
model II we will not give many details. Nevertheless it is useful 
to consider the supersymmetric stabilization points
\bea
U_{\rm crit}&=&\frac{1}{2 Z_{\rm crit}^{\frac{1}{2}}}
\sqrt{6 q_U +3 q_W - 3 \sqrt{4 q_U^2
+4 q_U q_W +9 q_W^2 -8 q_T^2}} \nn \\
T_{\rm crit}&=&
\frac{3q_T}{U}=\frac{6 q_T}
{Z_{\rm crit}^{\frac{1}{2}}\sqrt{6 q_U +3 q_W -3 \sqrt{4 q_U^2
+4 q_U q_W +9 q_W^2 -8 q_T^2}}}
\\
W_{\rm crit}&=&\frac{U^2 -6 q_W}{2 U}=
\frac{6 q_U -21 q_W- 3 \sqrt{4 q_U^2
+4 q_U q_W +9 q_W^2 -8 q_T^2}}
{4Z_{\rm crit}^{\frac{1}{2}}\sqrt{6 q_U +3 q_W- 3 \sqrt{4 q_U^2
+4 q_U q_W +9 q_W^2 -8 q_T^2}}}.
  \nn 
\eea
With some algebra we find that supersymmetric stabilization takes
place if
\be
q_U\geq\frac{3}{2}q_T\geq 0
\la{IIIb1}\ee
and
\be
q_W\leq -\frac{2}{9}\left(q_U-\sqrt{q_U^2-\frac{9}{4}q_T^2}\right).
\la{IIIb2}
\ee
If (\ref{IIIb1}) is saturated the critical point lies
directly on the $\ttt^2=T-\frac{1}{2} U -W=0$ (right) 
boundary, whereas 
saturation of (\ref{IIIb2}) implies stabilization on the
flop transition line $\ttt^3=W-U=0$. In fact, the stabilization
points of model II and model III agree  with each other 
for such a configuration
as predicted.
In both cases, the moduli metric and the gauge couplings
are finite. At the flop transition we expect $Z, \partial Z$ and $V$ to
be continuous. However, $V$ should have a kink, according to our
discussion in section 5.1. Figure 6 and 7 graphically illustrate
this behavior. The minima of $V$ and $Z^2$ are marked by an ellipse.

\begin{figure}
\centerline{\epsfxsize=130mm \epsfbox{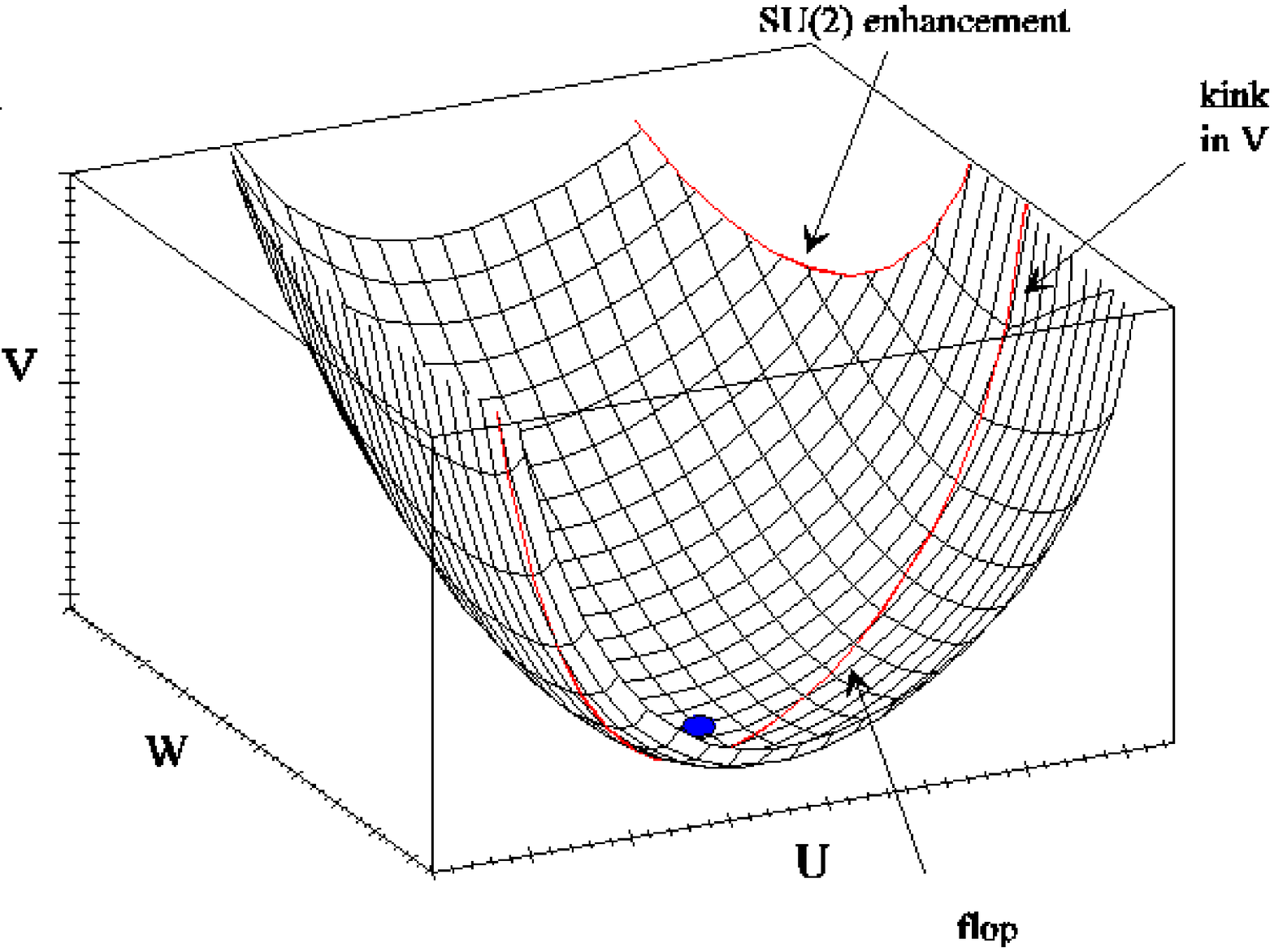}}
\medskip
\caption{Continuity and kink of $V$}
\bigskip

\centerline{\epsfxsize=130mm \epsfbox{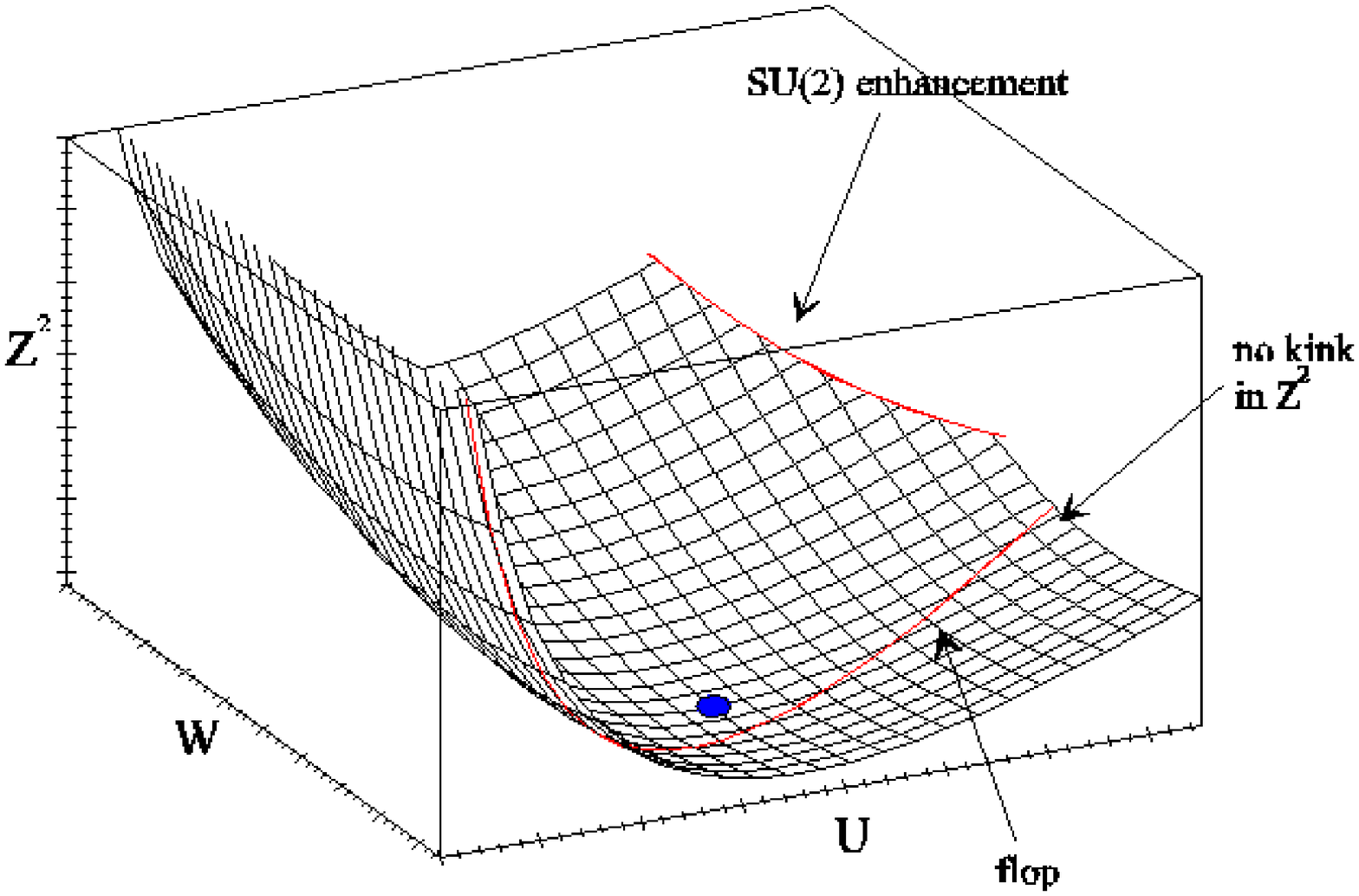}}
\medskip
\caption{Continuity and smoothness of $Z^2$}
\end{figure}

Figure 8  shows how the stabilization
point smoothly moves over the flop boundary if we vary $q_W$. 

\begin{figure}
\centerline{\epsfxsize=55mm\epsfbox{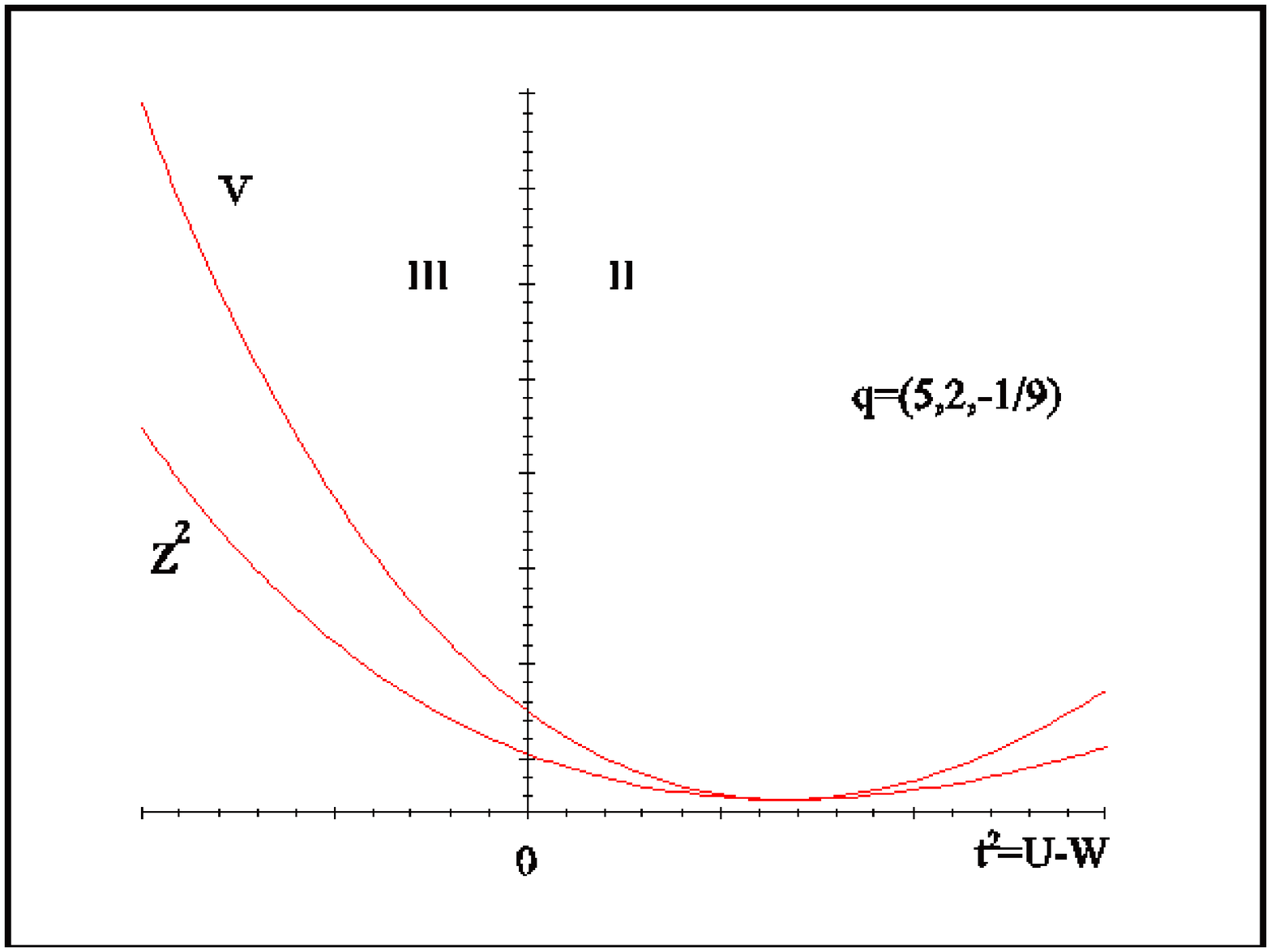}
\epsfxsize=55mm\epsfbox{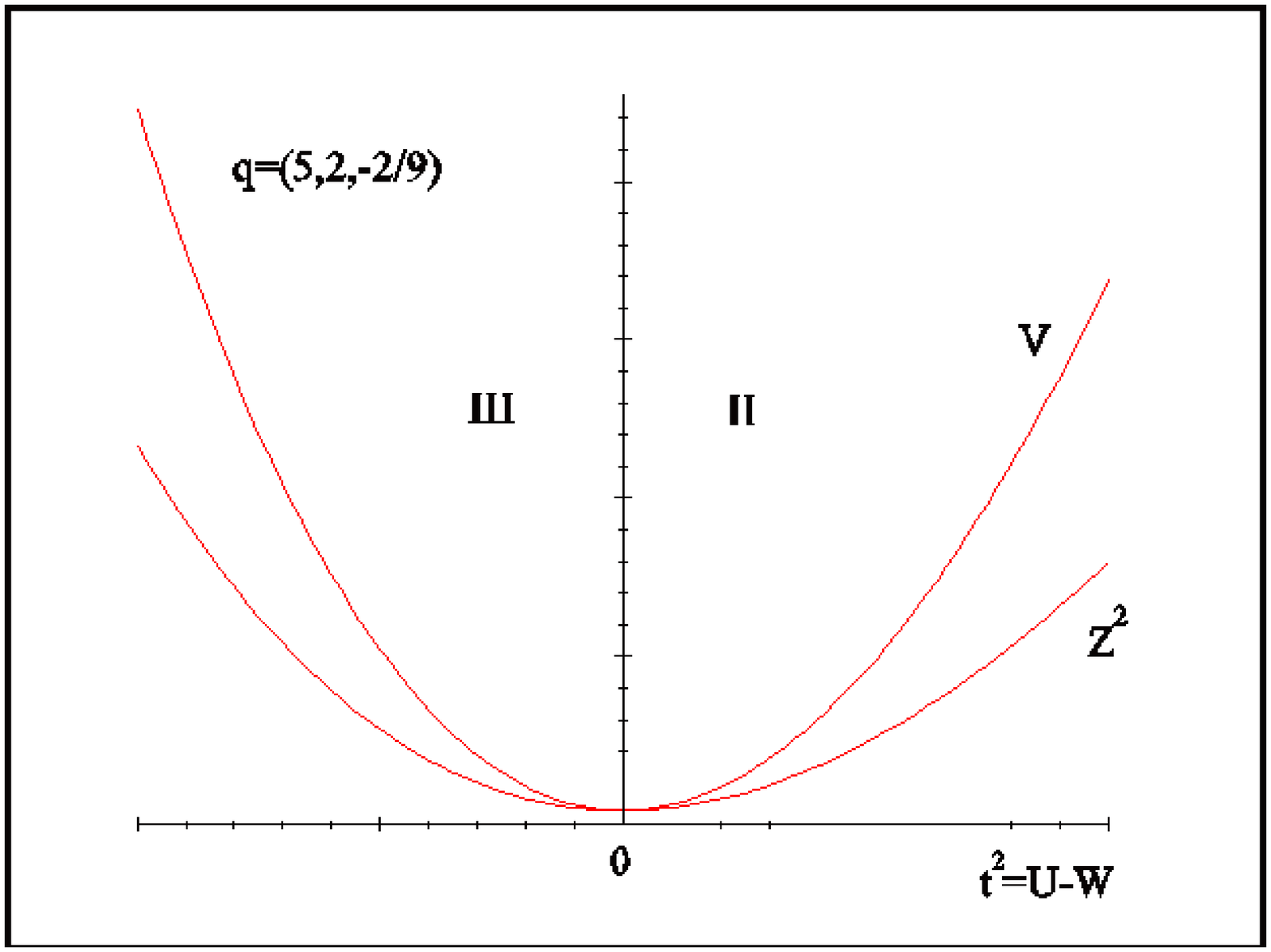}
\epsfxsize=55mm\epsfbox{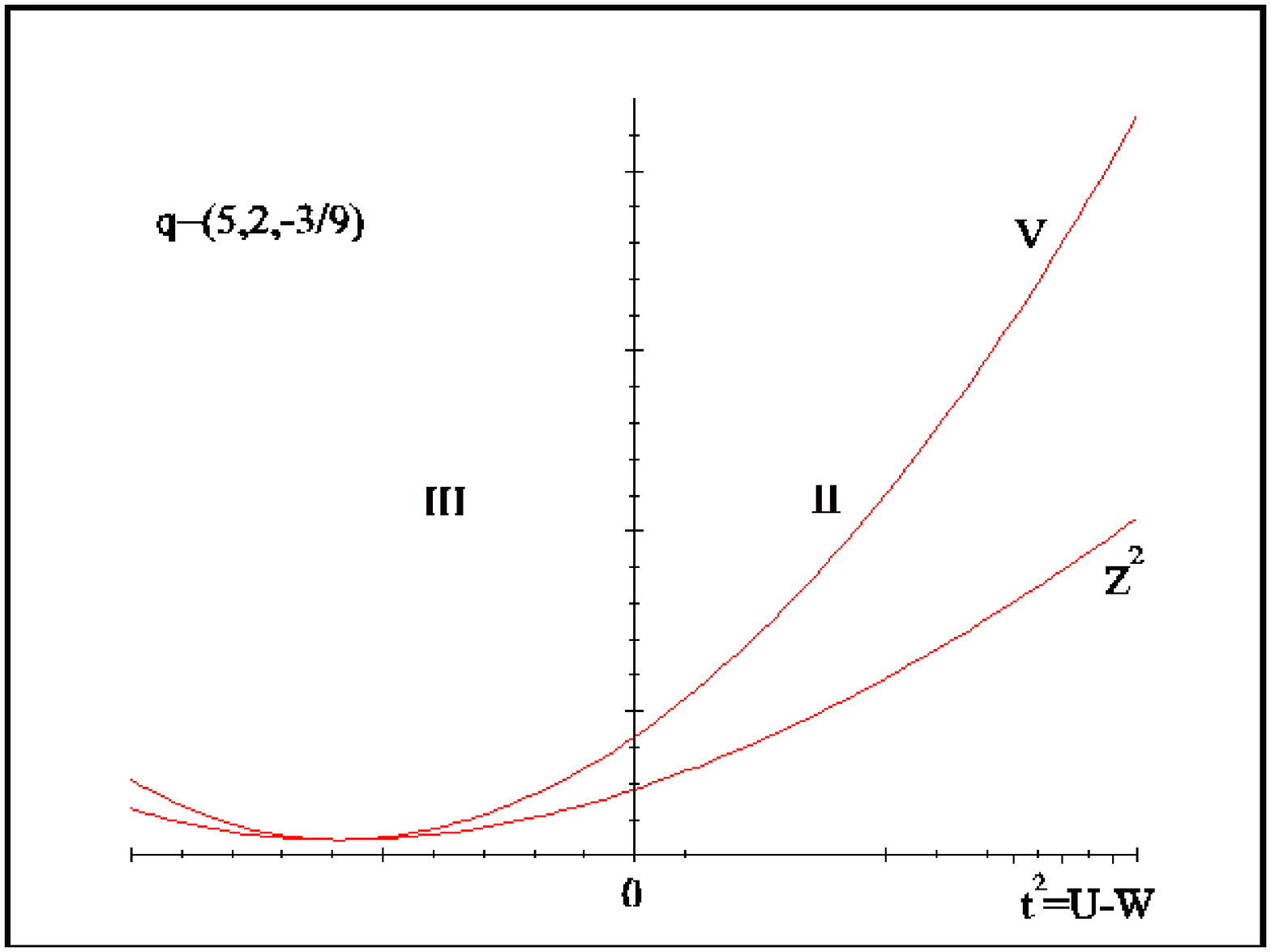}}
\medskip
\caption{Stabilization for configurations $(q_U,q_T,q_W)=
(5,2,-1/9),(5,2,-2/9),(5,2,-3/9)$}
\end{figure}

The right hand side boundary $\ttt^2=T-\frac{1}{2}U -W=0$ 
is different in the sense
that there is no flop transition to another Calabi-Yau space
possible, although the metric is also regular. Hence,
the conclusion is that this line corresponds to
an end of the moduli space in finite distance. Let us see
again whether strings become tensionless at this point.
Indeed, one finds that the magnetic central charge for 
string states with charges
\be
\mtt=\pm (1,-2,1)
\ee
vanishes. However, in this case $Z_m$ behaves as
\be
Z_m\sim \ttt^2
\ee
which implies that the tension of the string is decreasing
much slower than we found for the two non-flop boundaries of
  model II. 
This behavior again agrees with the 
one expected in \cite{wittenM} for the  case where a divisor
shrinks to a curve and symmetry enhancement to $SU(2)$ takes place.
This symmetry enhancement is indeed expected in our model, as
one can perform a redefinition of variables to write the prepotential
as $\Nu_{III} = S'T'U' -\frac{1}{3}U'^3$. In these variables 
$\ttt^2\rightarrow
0$ corresponds to $T'=U'$ which is known to be a symmetry enhancement
point.

At the other boundary $\ttt^1\rightarrow 0$ we again find
the phenomenon of strings becoming tensionless faster
than any particle  becomes massless.
We observe that at the
end of the moduli space of our example 
special geometry remains valid if
a divisor shrinks to a curve. If the divisor shrinks to
a point, special geometry breaks down and the moduli space metric
and gauge couplings degenerate or diverge.

\section{Conclusions}
In this paper we have studied phase transition of $d=5$ M-theory
compactifications
 on Calabi-Yau three-folds. We have probed the nature of the
phase transitions utilizing several physical potentials
including the electric BPS mass  $Z$, the magnetic string tension 
$Z_m$, the black hole potential $V$ and
the gauged central charge potential $P$. They are direct supersymmetric
and supergravity analogs of
the Landau-Ginzburg potentials used in thermodynamic systems that
undergo phase transitions. 

A lot of attention recently was directed to the properties of the entropy of 
the extreme black holes. The entropy is given by the area of the horizon and 
is related to the size of the infinite throat of $adS_2$.  It has a 
topological origin, being independent on the values of moduli far away from 
the horizon \cite{Ferraraextremal}. The value of the entropy is given by  
the central charge  extremized  in the moduli space \cite{Ferraraattractors}.
We found that in five dimensions the corresponding stabilization equations
take an extremely simple form.
 
For the extended objects $p>0$ other than black holes, which have a vanishing 
area of the horizon, the analogous critical behavior was not studied
 before. Here we have found that  the tension of the magnetic string
which is not equal to the area of the horizon,  but describes a size of an 
infinite throat of $adS_3$, is given by a critical value of the magnetic 
central charge.
We  found that the tension of the magnetic string in 5d can take  a 
minimal value $T^3 =  C_{IJK} m^I m^J m^K$ 
 for a given choice of winding numbers $m^I$ of an 11d fivebrane wrapped
on four-cycles inside the CY space. We expect that  
new results on critical behavior of various 
extended objects will emerge based on these results. 

We also
have shown explicitly that $V$ and $P$ admit new extrema which
are not related to supersymmetric stabilization.

One of the main goal in this work was to obtain a better understanding of
the structure of the  above supersymmetry 
potentials in the entire extended 
K\"ahler cone. In $d=5$, because of absence of non-geometric phases,
the extended K\"ahler cone is generically a geodesically incomplete,
bounded domain. Since $Z$ and $\partial Z$ turned out to be continuous
across flop transitions, and any critical point of
$Z$ must be a minimum, we were able to draw the important conclusion that
supersymmetric stabilization can take place in at most one point 
in the entire extended K\"ahler cone. The same conclusion was drawn for
the string tension $Z_m$. The analytic properties of $V$
were shown
to be slightly different. At the flop $V$ is continuous but not 
differentiable. It remains to be shown that also the critical
point of $V$ (for fixed charges) is unique. This result would
prove that the black hole entropy is unaffected even though a topology
change takes place as one moves away from the black hole horizon
to spatial infinity.

Finally we have studied in detail several explicit examples of Calabi-Yau
compactifications with  small  dimensions of the K\"ahler
moduli space.
 We  have
verified our general results of
previous sections on these examples 
and found that at the boundaries of the extended K\"ahler
cone, BPS magnetically charged strings become tensionless. In some cases,
these tensionless strings can be related to extremal phase transitions to
Calabi-Yau spaces with different Hodge numbers.

\section{Acknowledgments}

We are grateful for useful discussions with P. Candelas, S. Ferrara, 
E. Gimon, C. Johnson, X. de la 
Ossa, Y. Oz and E. Witten. The work of R.K., J.R., M.S. and
W.K.W. is supported by the NSF grant THY-9219345. The work of M.S. is also
supported by the 
Department of Energy under contract DOE-DE-FG05-91ER40627.
The work of S.-J. R. is supported in part by 
DOE DE-FG-02-90ER40542, NSF Grant PHY-9513835,  
NSF-KOSEF Bilateral Grant,
KOSEF Purpose-Oriented Research Grant and SRC Program, 
Ministry of Education
BSRI Program 97-2410 and the Monell Foundation and the Seoam Foundation
 Fellowships.

\newpage

%


\end{document}